\documentclass[structabstract]{aa}  
\usepackage{graphicx}
\usepackage{txfonts}
\usepackage{textcomp}
\usepackage{natbib}
\newcommand{\Ms}{{\ensuremath{\mathrm{M}_{\odot}}}}
\newcommand{\Ls}{{\ensuremath{\mathrm{L}_{\odot}}}}
\newcommand{\Rs}{{\ensuremath{\mathrm{R}_{\odot}}}}
\newcommand{\Teff}{{\ensuremath{T_{\rm eff}}}}
\newcommand{\Mpy}{\Ms\,{\rm yr}{\ensuremath{^{-1}}}}
\newcommand{\fitm}{{\rm fit}\ensuremath{_M}}

\newcommand{\dm}{\ensuremath{\dot M}}

\newcommand{\vsurf}{{\ensuremath{\mathrm{v_{surf}}}}}
\newcommand{\vcrit}{{\ensuremath{\mathrm{v_{crit}}}}}
\newcommand{\rapv}{{\ensuremath{\mathrm{\vsurf/\vcrit}}}}
\newcommand{\rapw}{{\ensuremath{\mathrm{\vsurf\over\vcrit}}}}
\newcommand{\jac}{{\ensuremath{j_\mathrm{accr}}}}

\newcommand{\jk}{{\ensuremath{j_\mathrm{K}}}}

\newcommand{\Mz}{{\ensuremath{M_\mathrm{ZAMS}}}}
\newcommand{\jcgs}{{\ensuremath{\mathrm{cm^2\,s^{-1}}}}}

\newcommand{\papI}{\citetalias{haemmerle2016a}}

\defcitealias{haemmerle2016a}{Paper~I}

\begin{document}

   \title{Massive star formation by accretion}
 \subtitle{II. Rotation: how to circumvent the angular momentum barrier?}

     \author{L. Haemmerl\'e\inst{\ref{inst1},\ref{inst2}},      
     P. Eggenberger\inst{\ref{inst1}},
         G. Meynet\inst{\ref{inst1}}, 
          A. Maeder\inst{\ref{inst1}},      
        C. Charbonnel\inst{\ref{inst1},\ref{inst4}},
        R. S. Klessen\inst{\ref{inst2},\ref{inst3}}
}

  \authorrunning{Haemmerl\'e et al.}

\institute{Observatoire de Gen\`eve, Universit\'e de Gen\`eve, chemin des Maillettes 51, CH-1290 Sauverny, Switzerland    \label{inst1}
\and Institut f\"ur Theoretische Astrophysik, Zentrum f\"ur Astronomie der Universit\"at Heidelberg,
Albert-Ueberle-Str. 2, D-69120 Heidelberg, Germany      \label{inst2}
\and Interdisziplin\"ares Zentrum f\"ur wissenschaftliches Rechnen der Universit\"at Heidelberg,
Im Neuenheimer Feld 205, D-69120 Heidelberg, Germany    \label{inst3}
\and IRAP, UMR 5277 CNRS, 14 Av. \'E. Belin, 31400 Toulouse, France     \label{inst4}}

   \date{Received 28 November 2016 ; accepted 2 March 2017}

 
\abstract
{Rotation plays a key role in the star-formation process, from pre-stellar cores to pre-main-sequence (PMS) objects.
Understanding the formation of massive stars requires taking into account the accretion of angular momentum during their PMS phase.}
{We study the PMS evolution of objects destined to become massive stars by accretion,
focusing on the links between the physical conditions of the environment and the rotational properties of young stars.
In particular, we look at the physical conditions that allow the production of massive stars by accretion.}
{We present PMS models computed with a new version of the Geneva Stellar Evolution code self-consistently including accretion and rotation
according to various accretion scenarios for mass and angular momentum.
We describe the internal distribution of angular momentum in PMS stars accreting at high rates
and we show how the various physical conditions impact their internal structures, evolutionary tracks, and rotation velocities
during the PMS and the early main sequence.}
{We find that the smooth angular momentum accretion considered in previous studies
leads to an angular momentum barrier and does not allow the formation of massive stars by accretion.
A braking mechanism is needed in order to circumvent this angular momentum barrier.
This mechanism has to be efficient enough to remove more than 2/3 of the angular momentum from the inner accretion disc.
Due to the weak efficiency of angular momentum transport by shear instability and meridional circulation during the accretion phase,
the internal rotation profiles of accreting stars reflect essentially the angular momentum accretion history.
As a consequence, careful choice of the angular momentum accretion history allows circumvention of any limitation in mass and velocity,
and production of stars of any mass and velocity compatible with structure equations.}
 {}
 
 
\maketitle
%

\section{Introduction}
\label{sec-intro}

Star formation requires mechanisms that extract angular momentum from collapsing cores.
If angular momentum was locally conserved during the collapse,
initial turbulent motions would lead to high rotation velocities as contraction proceeds,
exceeding the break-up velocity and preventing the collapse towards stellar densities \citep{spitzer1978,bodenheimer1995,maeder2009}.
Observed rotation velocities of T Tauri stars reveal angular momentum losses during the pre-main sequence (PMS) evolution
efficient enough to maintain the rotation period of contracting stars to a constant value
despite a decrease in the moment of inertia by several orders of magnitude \citep{bouvier1997}.
This fact is thought to result from the action of the convective dynamo of low-mass stars
that produces a magnetic field and couples contracting stars to their accretion discs at large distances \citep{konigl1991}.

Rotating models of PMS low-mass stars have been published by several authors
\citep[e.g.][]{pinsonneault1989,pinsonneault1990,bouvier1997,eggenberger2010,eggenberger2012,amard2016}.
The PMS evolution of low-mass stars is expected to proceed essentially through a canonical constant mass contraction
due to their long Kelvin-Helmholtz (KH) time \citep{hayashi1961b,larson1969,stahler1983}, meaning that accretion can be neglected.
In contrast, PMS evolution of intermediate- and high-mass stars is dominated by accretion \citep{bernasconi1996a,norberg2000,behrend2001},
at rates as high as $10^{-3}$ \Mpy\ \citep{wolfire1987}.
But models including accretion and rotation simultaneously and self-consistently are still lacking.

PMS models at high accretion rates have been computed by different authors, without the effects of rotation
\citep{omukai2001,omukai2003,yorke2008,hosokawa2009,hosokawa2010,hosokawa2013}.
According to these models, rapid accretion can significantly modify the stellar structure and the evolutionary track
compared to the low-rate case relevant for low-mass star formation, in particular by enhancing the swelling experienced by accreting PMS stars,
leading to radii larger than 100 \Rs.
This swelling corresponds to the increase in the stellar radius
produced by the internal luminosity wave (first described by \citealt{larson1972}),
i.e. the internal redistribution of entropy when the opacity in central regions decreases abruptly.
But the sensitivity of the swelling on the physical conditions makes PMS evolution with rapid accretion an open issue.
In \citet[hereafter \papI]{haemmerle2016a}, we showed that with accretion through a disc, the swelling can be significantly inhibited
by choosing initial conditions with a high entropy content.
We expect the rotational properties of PMS stars to differ significantly depending on the details of these evolutionary features.
Including rotation in accreting models could thus give useful constraints on the formation scenario of massive stars.

Recently, \cite{lee2016} studied the effect of rotation on the evolution of Population III stars that accrete at rates as high as $4\times10^{-3}$ \Mpy.
Indeed, at zero metallicity, the absence of dust and heavy elements makes the pre-stellar clouds hotter,
the Jeans masses larger, and the accretion rates higher.
At such high rates, the star approaches the Eddington limit during the swelling \citep{omukai2001,omukai2003}.
Moreover, with fast rotation, the centrifugal force lowers the effective Eddington limit; the so-called $\Omega\Gamma$-limit \citep{maeder2000}.
\cite{lee2016} obtained that the $\Omega\Gamma$-limit is reached at masses of 20 to 40 \Ms, preventing further accretion.
However, rotation in their models was not treated simultaneously with stellar evolution, but post-processed from non-rotating models
computed with the MESA code \citep{paxton2011,paxton2013,paxton2015}.
This procedure did not allow for inclusion of differential rotation, and the authors assumed solid-body rotation.
Moreover, they considered a unique class of angular momentum accretion histories,
taking in each model a constant fraction of the Keplerian angular momentum.
Interestingly, we notice that in their models, at $10^{-3}$ \Mpy\  the pre-MS swelling leads to radii lower than 20 \Rs,
which supports our result in \papI\ that a high initial entropy content can  drastically inhibit the swelling.

In \cite{haemmerle2013}, we presented models including accretion and rotation simultaneously.
However, these models were computed with an old version of the Geneva code, that contained an inconsistency in case of accretion.
The treatment of accretion in the Geneva code has recently been improved \citep{haemmerle2014}.
We described in \papI\ the effects of these improvements in the non-rotating case.
The present paper is devoted to the case with both accretion and rotation simultaneously.
These improvements allow us to present here the first PMS models for the formation of massive stars
that include accretion and rotation  simultaneously and self-consistently.
In Sect.~\ref{sec-code}, we describe the treatment of rotation in the code and the physical conditions assumed in our models.
Sect.~\ref{sec-mod} is devoted to the effect the various physical conditions have on PMS evolution.
In Sect.~\ref{sec-ms}, we present a birthline computed with the effect of rotation and
we look at the impact of PMS evolution with accretion of angular momentum on the MS evolution.
We discuss our results in Sect.~\ref{sec-bla} and summarise our conclusions in Sect.~\ref{sec-outro}.

\section{Rotation in the Geneva code with accretion}
\label{sec-code}

\subsection{Improvements in the code}
\label{sec-code-imp}

A full description of the treatment of accretion in the Geneva Stellar Evolution code
has already been given  in \papI\ in the non-rotating case.
We described the recent improvements to the code and how they modify the results
compared to the versions used in \cite{bernasconi1996a}, \cite{norberg2000}, \cite{behrend2001} and \cite{haemmerle2013}.
A detailed description of the treatment of rotation without accretion in the code is given in \cite{eggenberger2008}.
In the present section, we describe how the treatment of rotation with accretion is improved in the new version of the code
A full description of the simultaneous treatment of accretion and rotation in the new version of the code
is also available in \cite{haemmerle2014}.

The new improvements in the code are related to the time derivatives and the resulting time variation of the physical quantities.
Among the four equations of structure, the only one that refers to time is the equation of energy conservation, which reads
\begin{equation}
{dL_r\over dM_r}=\epsilon_n-T\,\left.{ds\over dt}\right\vert_{M_r},
\label{eq-e}\end{equation}
where $M_r$ is the mass enclosed in a shell of radius $r$, $L_r$ the luminosity at $r$, $\epsilon_n$ the energy generation rate by nuclear reactions,
$T$ the temperature, $s$ the specific entropy and $t$ the time.
Thus it is the equation that makes the star evolve.
As described in detail in \papI, in the new version of the code, we removed an inconsistency in this equation
that leads to artificial loss of entropy and to artificially low radii during the PMS swelling phase.
However, when we add rotation, we have to solve, in addition to the usual equations of stellar structure,
the equation of angular momentum transport.
In the Geneva code, rotation is treated with the assumption of \textit{shellular rotation} (\citealt{meynet1997}):
differential rotation is allowed in the radial direction, but we force the angular velocity to be uniform on each isobar.
The justification of this assumption is that horizontal turbulence that is produced by differential rotation
is efficient enough to maintain each isobar in solid-body rotation (\citealt{zahn1992}).
In addition, each convective region is also assumed to rotate as a solid body.
In radiative regions, two types of rotational instabilities are included: diffusion by shear and advection by meridional currents.
The equation of angular momentum transport reads
\begin{equation}
\rho\,{d\over dt}\left.\left(r^2\Omega\right)\right\vert_{M_r}={1\over5r^2}{d\over dr}\left(\rho r^4\Omega\,U(r)\right)
+{1\over r^2}{d\over dr}\left(\rho D(r)\,r^4{d\Omega\over dr}\right),
\label{eq-j1}\end{equation}
where $\Omega$ is the angular velocity, $U(r)$ the amplitude of the radial component of the meridional velocity,
and $D(r)$ the diffusion coefficient for shear instability (using the prescription of \citealt{maeder1997}).
We note that no magnetic field is taken into account.
In particular, no Tayler-Spruit dynamo is included.
However, we discuss the effect of an internal magnetic coupling in a semi-analytical way in Sect.~\ref{sec-bla-sam}.

As Eq.~(\ref{eq-e}), Eq.~(\ref{eq-j1}) refers to time through the Lagrangian derivative (of the angular momentum in this case)
\begin{equation}
{d\over dt}\left.\left(r^2\Omega\right)\right\vert_{M_r}={D\over Dt}\left(r^2\Omega\right).
\label{eq-dt}\end{equation}
The spatial coordinate used in the Geneva code is not, however, the Lagrangian coordinate $M_r$, but the relative mass coordinate
\begin{equation}
\mu={M_r\over M}.
\end{equation}
In the presence of accretion, $\mu$ is not a Lagrangian coordinate since $M$ increases with time.
With this coordinate, Eq.~(\ref{eq-j1}) becomes
\begin{eqnarray}
\rho\,{d\over dt}\left.\left(r^2\Omega\right)\right\vert_\mu-\rho\mu\,{\dot M\over M}{d\over d\mu}\left(r^2\Omega\right)
={1\over5r^2}{d\over dr}\left(\rho r^4\Omega\,U(r)\right)
\\\qquad\qquad\qquad\qquad\qquad\qquad+{1\over r^2}{d\over dr}\left(\rho D(r)\,r^4{d\Omega\over dr}\right).
\label{eq-j2}\end{eqnarray}
Eq.~(\ref{eq-j2}) contains an additional term compared to Eq.~(\ref{eq-j1}),
which is proportional to the accretion rate and the internal gradient of angular momentum.
Since, in general, the angular momentum increases outwards and $\dot M>0$ during accretion, this additional term is negative.
In the previous version of the code, the effect of this additional term was neglected,
leading to artificial loss of angular momentum.
Indeed, if we neglect this negative term, the term in $d(r^2\Omega)/dt$ has to be smaller.
The effects of these improvements on the models are illustrated in Appendix~\ref{app-imp}.

\subsection{Angular momentum accretion rate}
\label{sec-code-dj}

When we add rotation in the context of accretion, we also have to specify, in addition to the mass accretion rate,
the angular momentum accretion rate
\begin{equation}
\dot J={dJ\over dt}={dJ\over dM}{dM\over dt}=:\jac\times\dm
\label{eq-dj},\end{equation}
where \jac\ is the specific angular momentum of the material that is accreted.

The procedure applied in order to accrete angular momentum in the Geneva code is the following.
When we accrete mass during a time-step $dt$, we add a new layer in the outer boundary of the stellar interior,
defined as the region where $M_r/M<\fitm$ for a given value of \fitm\ (typically $\fitm=0.999$).
In the envelope ($M_r/M>\fitm$), several simplifying assumptions are made, for reasons of numerical stability.
In addition to assuming that $dL_r/dM_r=0$, we assume in the rotating case that the envelope rotates as a solid body,
with the same angular velocity as the external layer of the interior.
The new layer added between the interior and the envelope is defined as having exactly the mass accreted during $dt$, that is, $dM=\dot M\,dt$.
Once this layer has been added, we have to define its physical properties.
The choice of the thermal properties of this new layer has already been described in \papI,
and corresponds to the assumption of {\it cold disc accretion}.
In the rotating case, we also have to specify the angular velocity $\Omega_{\rm accr}$ of this new layer,
that we choose to be
\begin{equation}
{2\over3}\,r_{\rm fitm}^2\times\Omega_{\rm accr}=\jac={\dot J\over\dot M},
\label{eq-omac}\end{equation}
where $r_{\rm fitm}$ is the radius corresponding to $M_r/M=\fitm$.
The factor 2/3 in Eq.~\ref{eq-omac} reflects the fact that the material is distributed over a spherical shell.

In the present work, we use various prescriptions for $\dot J$.
The first is the \textit{smooth angular momentum accretion} (\textquoteleft smooth-J\textquoteright\ hereafter),
which states that the angular velocity of the material that is accreted is the same as that of the stellar surface before the material is added.
This prescription was used in \cite{haemmerle2013}.
We note that in this case, the only free parameter for rotation is the rotation velocity of the initial model.
In the second case considered here, we aim to fix a constant value of $\jac=dJ/dM$, that is, the specific angular momentum in the accreted material.
This gives an angular momentum accretion rate that is proportional to the mass accretion rate from Eq.~(\ref{eq-dj}).
However, the accretion of angular momentum in the Geneva code cannot be controlled as freely as the accretion of mass
in the cases where angular momentum transport is significant in the external layers of the star,
for instance when the external layers are convective, which implies instantaneous angular momentum redistribution.
In Sect.~\ref{sec-ms}, we use a slightly different prescription, with $\jac\propto M^{0.8}$.
Finally, in Appendix~\ref{app-kep}, we discuss the case of angular momentum accretion according to a constant fraction of the Keplerian value.

The physical justifications of these angular momentum accretion laws are related to the complex mechanisms
that govern the inner part of accretion discs, such as magnetic fields, viscosity, or gravitational instability.
The role of these mechanisms in regulating the accretion of angular momentum is still far from being fully understood.
Observations of discs around massive young stellar objects (MYSOs)
are in good agreement with Keplerian rotation profiles \citep{cesaroni2005,kraus2010,keto2010c,ilee2016},
but the inner part of these discs are out of the reach of current observations.
At the stellar surface, the Keplerian velocity corresponds to the critical limit beyond which accretion from the disc onto the star cannot occur.
The critical velocity is defined as the rotation velocity at the stellar surface at which the centrifugal force cancels gravity
(i.e. $\vcrit=\sqrt{GM/R_{\rm eq,crit}}$) and corresponds to the upper limit compatible with hydrostatic equilibrium.
Thus a braking mechanism is expected in the inner region of the discs, probably related to the magnetic field
\citep{pudritz2007,hennebelle2009,joos2012,seifried2013,seifried2015},
the disc viscosity \citep{vorobyov2009,stacy2011}, or the gravitational torques \citep{jappsen2004,vorobyov2007}.
The magnetic field of massive MS stars is still an uncertain issue given the absence of a surface convective dynamo
and the small number of magnetic fields detected at the surface of observed massive stars \citep[see~e.g.][]{grunhut2016}.
The common existence of outflow jets around MYSOs accreting at high rates \citep{arce2007,dougados2009,reiter2013,reiter2014,reiter2015a,reiter2015b}
indicates that magnetic fields play a significant role in the accretion process for massive star formation
\citep{guzman2010,whelan2010,ellerbroek2014}.
Such outflow jets are not necessarily due to stellar magnetic fields, and could result from magnetised disc winds \citep{pudritz2005,pudritz2007}
or a superposition of aligned outflows from many low-mass stars that usually form alongside massive stars \citep{peters2014}.
But, whatever the mechanism that launches the jets, the fraction of angular momentum carried away
can eventually represent a significant fraction of the Keplerian value \citep{bodenheimer1995,pudritz2009a}.
Additionally, PMS objects that are destined to become massive stars by accretion
have a large convective envelope during their early accretion phase (see e.g. \papI), which could potentially produce a magnetic field.
The geometry of the accretion flow close to the stellar surface depends sensitively on the geometry of the magnetic field,
which is poorly constrained \citep{pudritz2007,romanova2015}.
Whether the intersections between the field lines and the stellar surface are concentrated near the poles or distributed isotropically,
the amount of angular momentum that is accreted will be significantly modified.
The exact determination of the angular momentum accretion history for realistic geometries of the magnetosphere
goes far beyond the scope of the present work, and here we only consider arbitrary assumptions
in order to explore the various possible scenarios.
The smooth-J accretion corresponds to the case where the rotation of the star is imposed to the inner parts of the accretion disc,
but assuming that this inner region does not influence the rotation of the star.
As we will see, this accretion law leads the rotation of the star to become critical when, typically, masses of approximately 8 \Ms\ are reached.
This clearly indicates that a coupling between the star and the inner disc must operate preventing the star from reaching the critical limit.
The constant-\jac\ accretion corresponds to the case where the accreted material is not influenced by the rotation velocity of the stellar surface,
and where any braking occurs upstream in the accretion flow.

\section{Impact of the physical conditions on PMS evolution with accretion and rotation}
\label{sec-mod}

\subsection{Smooth-J accretion: model at low mass-accretion rate}
\label{sec-mod-1}

We first describe a test model computed with the new version of the code at the low constant mass-accretion rate of
\begin{equation}
\dm=10^{-5}\,\Mpy,
\label{eq-mod1-dm}\end{equation}
assuming smooth-J accretion, as in \cite{haemmerle2013}.
Initial conditions are given by
\begin{equation}
M=0.7\,\Ms              \qquad  L=9.57\,\Ls     \qquad  \Teff=4130\rm\,K
\label{eq-mod1-ini},\end{equation}
with solar composition\footnote{\ 
   As in \papI: $Z=0.014$, with the abundances of \cite{asplund2005} and \cite{cunha2006},
   and a deuterium mass fraction of $5\times10^{-5}$ \citep{bernasconi1996a,norberg2000,behrend2001,haemmerle2013}.}.
This initial model has a radius of 6.06 \Rs\ and is fully convective.
Since solid-body rotation is assumed in convective regions, a unique value of the angular velocity defines the whole rotational inputs, here
\begin{equation}\begin{array}{l}
\Omega_{\rm ini}=10^{-7}\rm\,s^{-1}\\
\vsurf=0.42\rm\,km\,s^{-1}=0.28\%\,\vcrit.
\end{array}\label{eq-mod1-vini}\end{equation}
Such velocity is extremely low for a PMS star of 0.7 \Ms\ (smaller than 1\% of the critical velocity, Eq.~\ref{eq-mod1-vini}).
The reason for this choice is presented below.
For the moment, we only notice that the models have been computed for almost the whole PMS phase.
For the considered mass-accretion rate, this corresponds to the low- and intermediate-mass range: $M\lesssim8$ \Ms.

\begin{figure}\begin{center}
\includegraphics[width=0.45\textwidth]{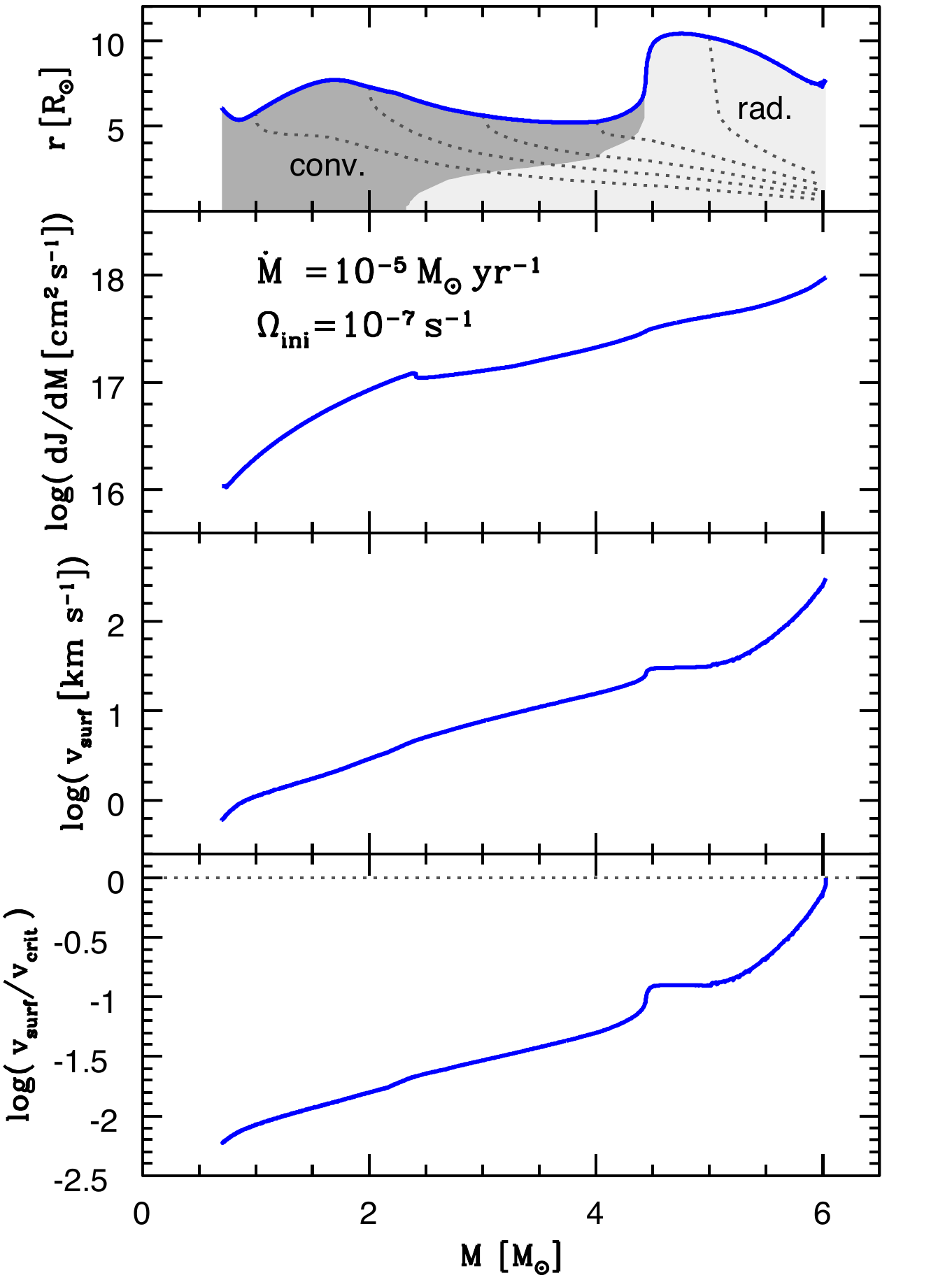}
\caption{Evolution of the internal structure, the angular momentum accretion rate, and the equatorial surface velocity
of the test-model at low mass-accretion rate described in Sect.~\ref{sec-mod-1},
as a function of the increasing stellar mass, which is a time-coordinate in the case of accretion.
The upper panel shows the radius (blue curve) and the internal structure,
with convective zones in dark grey and radiative zones in light grey.
The dark grey dotted curves are iso-mass of $M_r=1$, 2, 3, ... \Ms.
The second panel shows the angular momentum accretion rate in logarithm of $\jac=dJ/dM=\dot J/\dm$.
The third panel shows the surface velocity (at the equator) and the fourth panel shows the ratio of the surface velocity to the critical velocity.
(The critical limit is indicated by the horizontal grey dotted line.)}
\label{fig-mod1}
\end{center}\end{figure}

The evolution of the internal structure of this model is shown in the upper panel of Fig.~\ref{fig-mod1}.
At such a low rotation velocity, the effect of rotation on the evolutionary tracks (not shown here) and on the internal structure is negligible,
as with the original version of the code (see \citealt{haemmerle2013}).
A detailed description of the evolution of the internal structure for such a model can be found in \papI; here we summarise the main features.
The evolution starts with a fully convective structure.
The internal temperature is initially too low for nuclear burning, and the star takes its energy from gravitational contraction.
The radius decreases and the central temperature grows, until it exceeds $\sim10^6$ K (at $M\simeq1$ \Ms).
Following this, D-burning starts in the centre, the temperature stops growing, and the radius increases ($T_c\propto M/R$).
When D is exhausted, the star has to contract again, the radius decreases, and the central temperature grows again.
As a result, the opacity decreases, and at $M=2.4$ \Ms\ , a radiative core appears and grows in mass.
The high luminosity emerging from this hot radiative core is absorbed by the cold convective envelope, which expands
(\textit{luminosity wave}, \citealt{larson1972,hosokawa2010});
the star experiences a rapid swelling, that leads in this case to a maximum radius of 10.4 \Rs.
At that point, the convective envelope has disappeared and the fully radiative star goes through a KH contraction.

The second panel of Fig.~\ref{fig-mod1} shows the evolution of $\jac=dJ/dM=\dot J/\dot M$ (Eq.~\ref{eq-dj}),
that is, the angular momentum accretion rate given by the assumption of smooth-J accretion.
The value of \jac\ increases as the stellar mass grows, by two orders of magnitude between 0.7 and 6 \Ms.
The surface velocity at the equator (third panel) increases by almost three orders of magnitude in the same interval.
We notice that the increase becomes particularly strong for $M>5$ \Ms,
which corresponds to the phase of post-swelling KH contraction (upper panel),
where the external layers contract rapidly (see the iso-mass of 5 \Ms).

The lower panel of Fig.~\ref{fig-mod1} shows the evolution of the ratio \rapv.
The model starts with a surface velocity smaller than 1\% of the critical value, following the assumptions described in Eq.~\ref{eq-mod1-vini}.
Then, due to the increase of the stellar mass by one order of magnitude during the PMS,
the critical velocity increases slightly.
However, since the surface velocity increases by three orders of magnitude during the same phase,
the ratio \rapv\ increases by more than two orders of magnitude.
As a consequence, in this model, the stellar surface reaches the critical velocity at a mass of
\begin{equation}
M_{\rm max}=6\ \Ms.
\label{eq-mod1-mmax}\end{equation}
The critical velocity is the maximum value of the surface rotation velocity for a star in hydrostatic equilibrium;
if the surface velocity increases above this value, the centrifugal force dominates over gravity,
accretion stops, and the star begins to lose mass.
As a consequence, masses higher than this $M_{\rm max}$ cannot be reached in this scenario,
that is, the star cannot enter the range of massive stars.
In the present models, we did not include any mass-loss: we simply stopped the computation when $\vsurf=\vcrit$.

\begin{figure}\begin{center}
\includegraphics[width=0.45\textwidth]{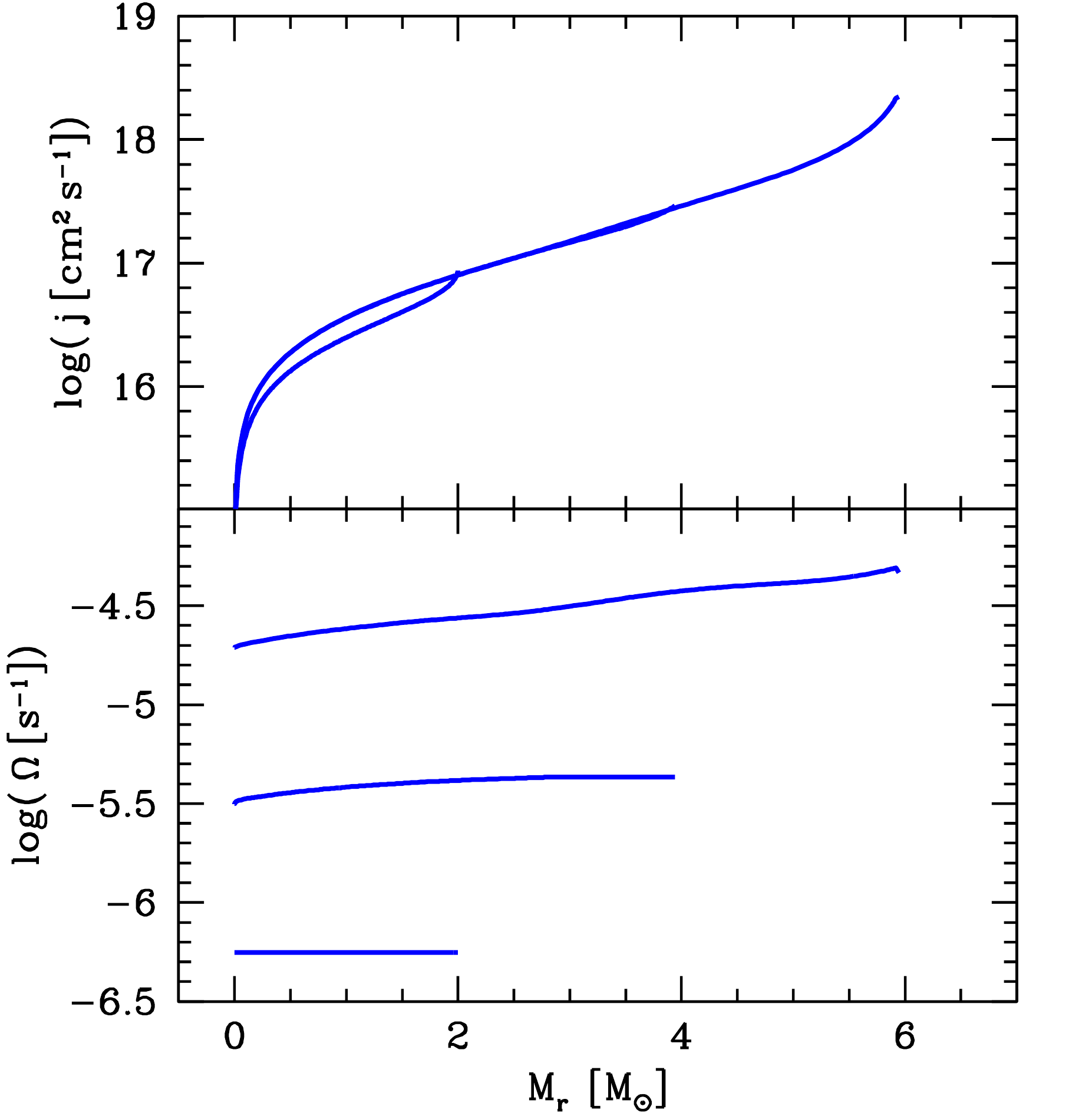}
\caption{Internal rotation profiles of the model at low mass-accretion rate described in Sect.~\ref{sec-mod-1}
for evolutionary stages $M=2$, 4, and 6 \Ms, as a function of the Lagrangian coordinate $M_r$.
The upper panel shows the profiles of the specific angular momentum $j={2\over3}\,r^2\,\Omega$
while the lower panel shows the angular velocity $\Omega$.}
\label{fig-mod1prof}
\end{center}\end{figure}

In order to understand the evolution of the surface rotation velocity,
we look at the internal rotation profiles along the evolution, when the model reaches stellar masses of 2, 4, and 6 \Ms\ (Fig.~\ref{fig-mod1prof}).
As is visible in the upper panel of Fig.~\ref{fig-mod1}, the star is still fully convective at $M=2$ \Ms.
At 4 \Ms\ , the radiative core has grown to $M_r=3\,\Ms$, so that the convective envelope still covers 25\% of the stellar mass.
At 6 \Ms, the star is fully radiative.
The upper panel of Fig.~\ref{fig-mod1prof} shows the profiles of the specific angular momentum on each shell of radius $r$, explicitly:
\begin{equation}
j={2\over3}\,r\times {\rm v_{eq}}(r)={2\over3}\,r^2\times\Omega,
\label{eq-j}\end{equation}
where $\rm v_{eq}(r)$ is the equatorial velocity of the shell of radius $r$.
Except in convective regions, the $j$-profiles match one another.
This means that angular momentum is locally conserved in radiative regions,
that is, angular momentum transport by shears and meridional circulation is negligible in the short timescale of the accretion phase.
The profiles at 2 and 4 \Ms\ differ due to the assumption of solid-body rotation in convective regions, implying instantaneous angular momentum redistribution in cases of non-homologous contraction.
The profiles at 4 and 6 \Ms\ differ only slightly in the layers with $3\,\Ms<M_r<4\,\Ms$,
corresponding to the convective envelope in the 4 \Ms\ profile.
Given the $j$-profiles, the $\Omega$-profiles (Fig.~\ref{fig-mod1prof}) depend solely on the relative contraction of each layer.
Solid rotation in convective zones is reflected in a flat $\Omega$-profile at $M=2\,\Ms$,
as well as in the layers with $M_r>3\,\Ms$ of the profile at $M=4\,\Ms$.
In radiative regions, the rapid contraction of the external layers (Fig.~\ref{fig-mod1}, upper panel)
leads to $\Omega$-profiles that increase outwards ($\dot\Omega/\Omega=-2\,\dot r/r$).
In particular, during the post-swelling contraction, the external layers contract in a very short time
(iso-mass of 5 \Ms\ on the upper panel of Fig.~\ref{fig-mod1}), leading to high rotation velocities (notice the logarithmic scale).
This is the reason why the surface velocity reaches the critical value during this stage.

\subsection{Varying the initial rotation velocity}
\label{sec-mod-vini}

As mentioned in Sect.~\ref{sec-code}, in the case of smooth-J accretion,
the value of the initial rotation velocity fixes the whole rotational history of the star.
In Sect.~\ref{sec-mod-1}, we showed that starting from a low rotation velocity,
the critical limit is reached before the star enters the range of massive stars.
Now, we try to decrease the initial rotation velocity even more, by one order of magnitude,
and we consider a model with
\begin{equation}\begin{array}{l}
\Omega_{\rm ini}=10^{-8}\rm\,s^{-1}\\
\rm\vsurf=0.042\,km\,s^{-1}=0.028\%\,\vcrit.
\end{array}\label{eq-vini59}\end{equation}
Except for the rotation velocity, we consider the same initial model as in Sect.~\ref{sec-mod-1} (Eq.~\ref{eq-mod1-ini}),
and we compute a birthline with the new version of the code, again at $10^{-5}$ \Mpy.

\begin{figure}\begin{center}
\includegraphics[width=0.45\textwidth]{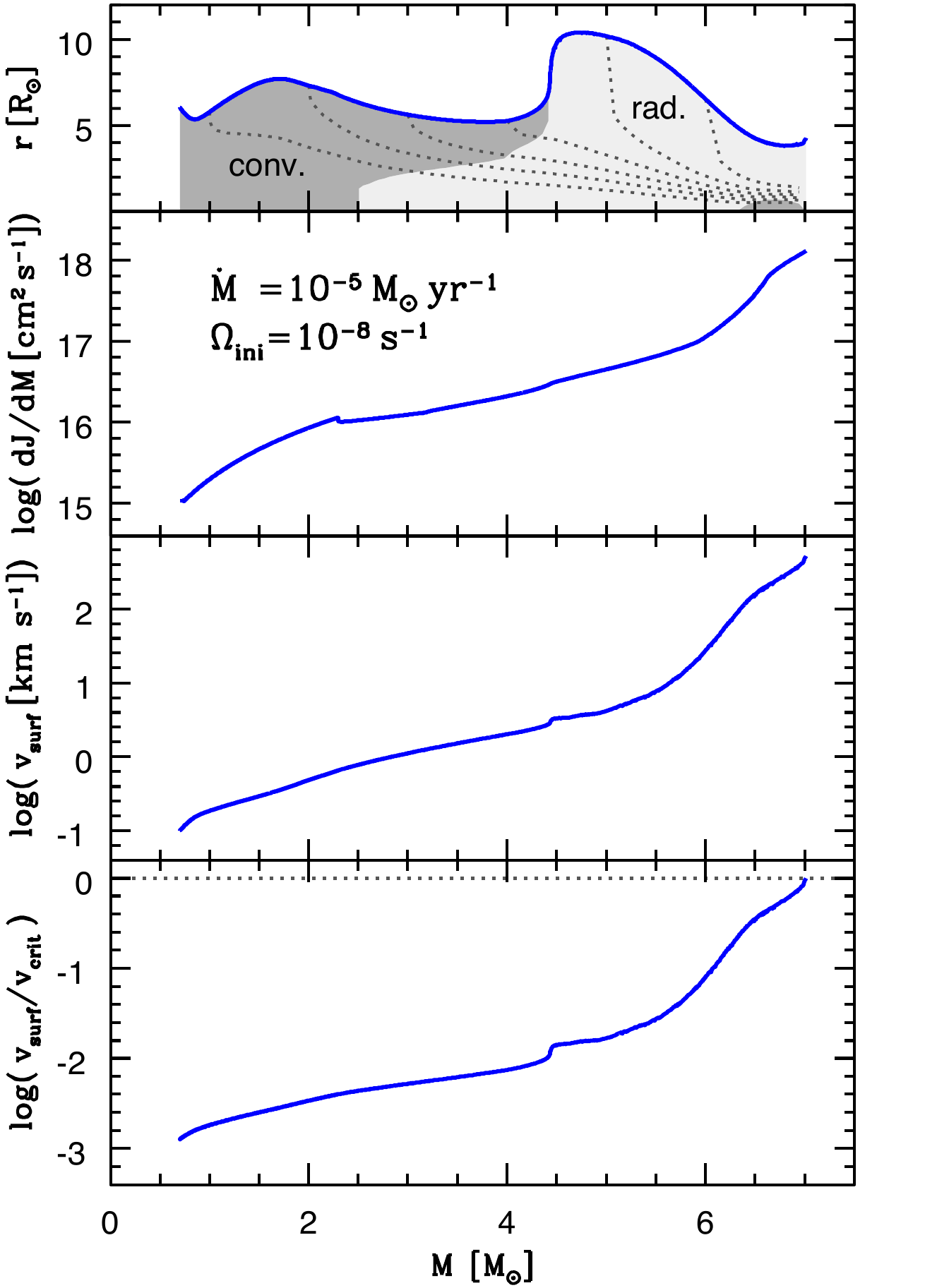}
\caption{As Fig.~\ref{fig-mod1}, for the slower rotator described in Sect.~\ref{sec-mod-vini}.}
\label{fig-vini}
\end{center}\end{figure}

The evolution of the surface rotation velocity of this model is shown in Fig.~\ref{fig-vini}.
As for the faster rotator described in Sect.~\ref{sec-mod-1},
the surface velocity increases by several orders of magnitude along the intermediate-mass range.
Again, the increase is particularly strong during the post-swelling contraction ($M\gtrsim5$ \Ms).
In this case, due to the low initial velocity, the stellar surface reaches the critical velocity
at a slightly higher mass than for the faster rotators described in Sect.~\ref{sec-mod-1}.
Here the maximum mass that can be accreted is
\begin{equation}
M_{\rm max}=7\,\Ms.
\label{eq-vini-mmax}\end{equation}
Although this value is higher than that of the previous case, the difference remains relatively small.
Indeed, we see that when we decrease $\Omega_{\rm ini}$ by one order of magnitude,
we increase $M_{\rm max}$ by 17\% only.
As a consequence, even with this lower initial velocity, the star cannot enter the high-mass range.

\subsection{Varying the mass accretion rate}
\label{sec-mod-dm}

We now consider various accretion rates
in order to see how the choice of the accretion history can affect the internal structure of the star and its final mass.
The accretion rate used in Sect.~\ref{sec-mod-1} is lower than what is expected for massive star formation,
and we consider, here, more appropriate rates, such as a constant rate of $10^{-3}$ \Mpy\ , and the Churchwell-Henning (CH) accretion law.

The CH law \citep{behrend2001} has been described in detail in \papI, where we have shown that
it reproduces well the upper envelope of Herbig Ae/Be stars on the Hertzsprung-Russell (HR) diagram.
This law depends on a parameter $f$, which reflects the fraction of the mass from the accretion flow that is effectively accreted by the star,
the rest being ejected through bipolar outflows.
As shown in \papI, the best fit for intermediate- and high-mass stars is obtained with $f=1/11$, and thus we use this value in the present section.
We recall here that this rate is time dependent, increasing with the evolution, from $\sim10^{-5}$ \Mpy\
in the low- and intermediate-mass range to $\sim10^{-3}$ \Mpy\ in the high-mass range.

Numerical convergence is difficult for low-mass models that accrete at rates as high as $10^{-3}$ \Mpy.
Thus we started the computation of the model with constant high rate at
\begin{equation}
M=2\,\Ms                \qquad  L=124\,\Ls\     \qquad  \Teff=4272\rm\,K.
\label{eq-ini2}\end{equation}
This initial model has a radius of 20.5 \Rs\ and is fully convective.
In the case of CH accretion, the early accretion rate is still low
and we use the same initial model as in the previous sections (Eq.~\ref{eq-mod1-ini}).
In both cases, we consider $\Omega_{\rm ini}=10^{-8}\rm\,s^{-1}$.

\begin{figure}\begin{center}
\includegraphics[width=0.45\textwidth]{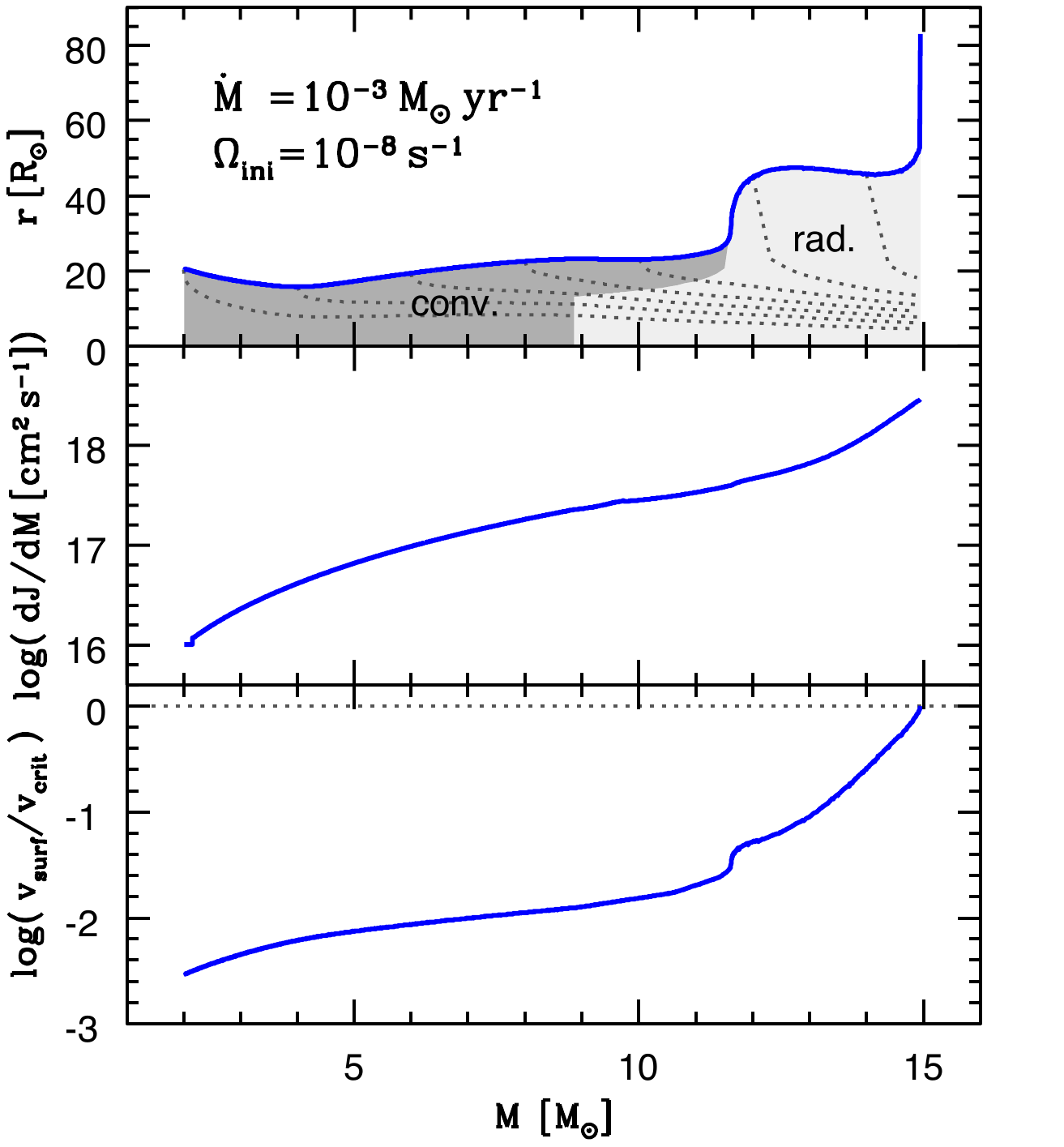}
\caption{As in Fig.~\ref{fig-mod1} (without \vsurf), for the model with $\dm=10^{-3}$ \Mpy\ and $\rm\Omega_{ini}=10^{-8}\,s^{-1}$,
described in Sect.~\ref{sec-mod-dm}.
The iso-mass corresponds to $M_r=2$, 4, 6, ... \Ms.}
\label{fig-dm3}\end{center}\end{figure}

\begin{figure}\begin{center}
\includegraphics[width=0.45\textwidth]{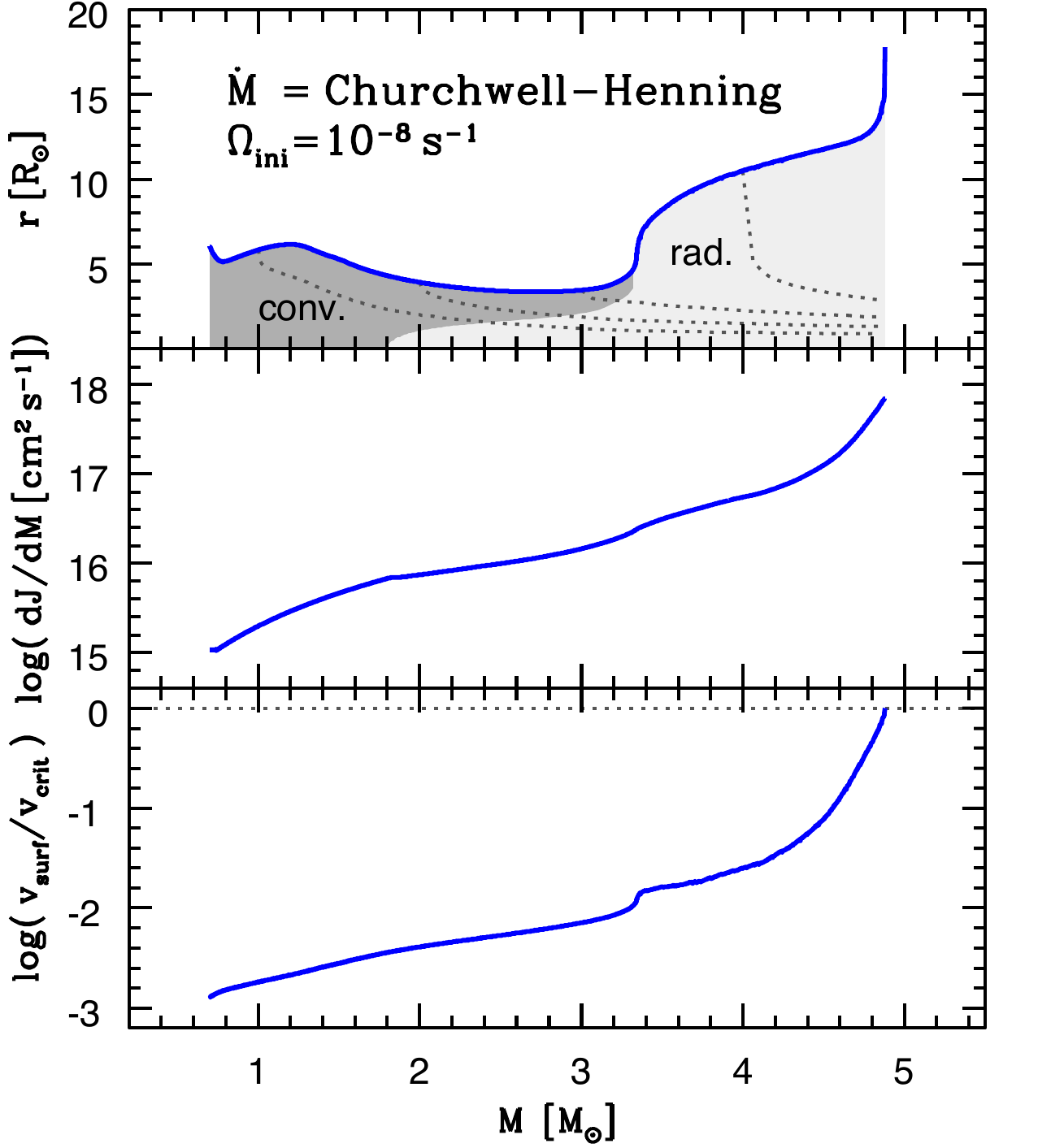}
\caption{As in Fig.~\ref{fig-mod1} (without \vsurf), for the model with CH accretion and $\rm\Omega_{ini}=10^{-8}\,s^{-1}$,
described in Sect.~\ref{sec-mod-dm}.}
\label{fig-dmch} \end{center}\end{figure}

The evolution of the ratio \rapv\ for these two rates is shown in Figs.~\ref{fig-dm3} and \ref{fig-dmch}.
In both cases, the surface velocity and the specific angular momentum accreted increase monotonically,
with a particularly abrupt jump during the swelling phase, when the convective envelope disappears,
and the external layers of the fully radiative star contract rapidly.
It leads to the critical limit, ending the accretion process at
\begin{equation}
M_{\rm max}=15\,\Ms     \qquad  M_{\rm max}=4.9\,\Ms
\label{eq-dm-mmax},\end{equation}
for $10^{-3}$ \Mpy\ and CH accretion, respectively.
Notice that when the surface velocity approaches the critical limit, the effect of rotation on the stellar structure becomes significant
by increasing the radius (by nearly a factor 2 in the $\dm=10^{-3}\,\Mpy$ case).
However, in the present models, such a phase is extremely short-lived, and moreover the effect concerns only the very external layers.
We see here that the high constant accretion rate allows the star to enter the high-mass range only marginally,
while the CH rate leads to a stronger limitation than in the cases considered previously.
In particular, the increase in the accretion rate by two orders of magnitude (from $10^{-5}$ to $10^{-3}$ \Mpy)
produces an increase in $M_{\rm max}$ by less than a factor 2.

\subsection{Varying the angular momentum accretion history}
\label{sec-mod-dj}

\begin{figure}\begin{center}
\includegraphics[width=0.45\textwidth]{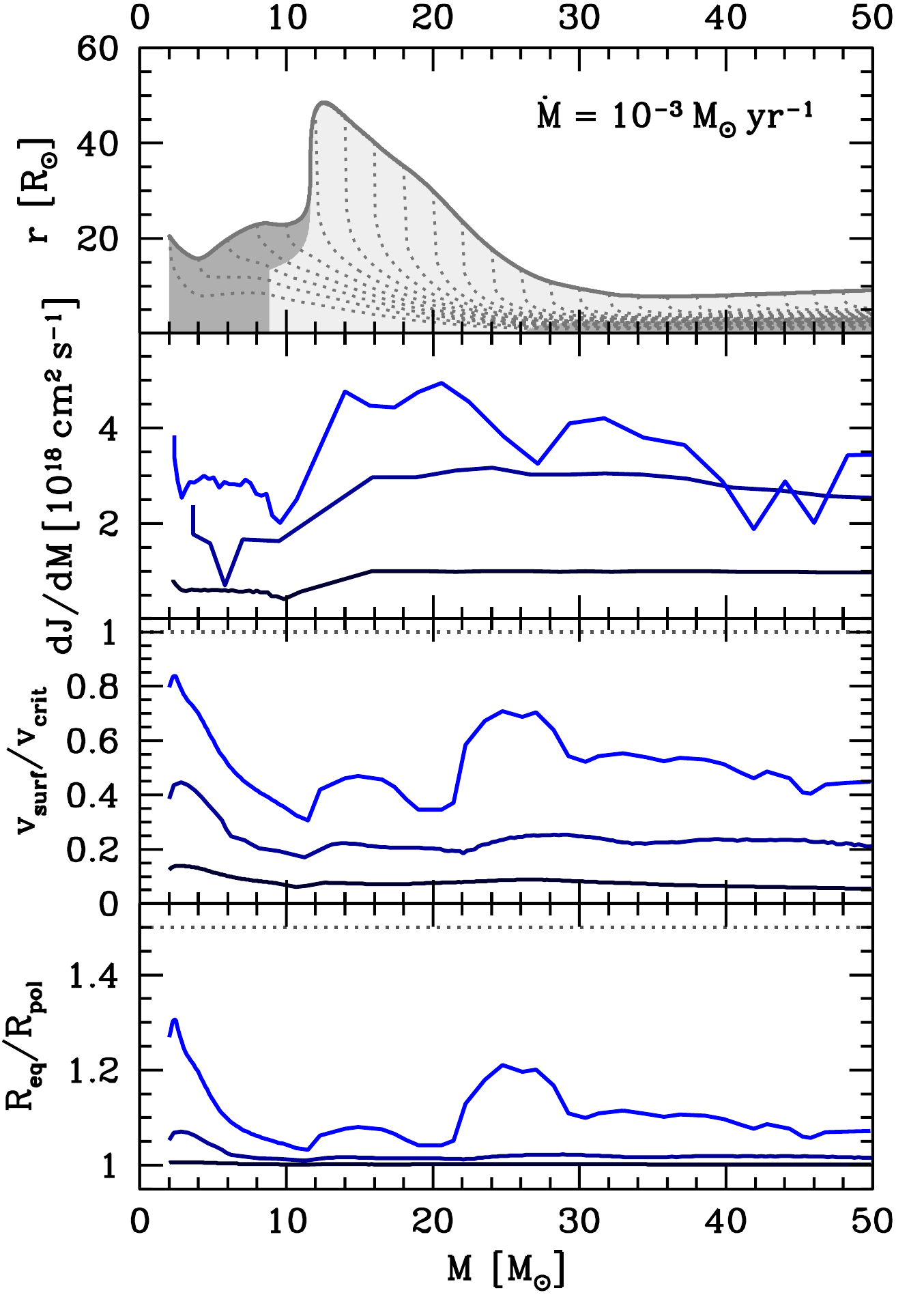}
\caption{Varying the angular momentum accretion history, for $\dm=10^{-3}$ \Mpy.
The upper panel shows the internal structure of the non-rotating model:
the upper curve is the stellar radius, the dark grey regions are convective, the light grey is radiative,
and the dotted lines are ios-mass of 2, 4, 6, 8, ... \Ms.
The second panel shows the three J-accretion histories considered and
the third one, specifically, shows the evolution of \rapv\ for the corresponding models (the grey dotted horizontal line indicates the critical limit),
and the bottom one shows the deformation of the star due to rotation ($R_{\rm eq}$ and $R_{\rm pol}$ are the equatorial and polar radii, respectively).}
\label{fig-dj}
\end{center}\end{figure}

Then we try to modify the J-accretion history by fixing a constant specific angular momentum in the accreted material ($dJ/dM=$\,cst).
However, as mentioned in Sect.~\ref{sec-code-dj}, accretion of angular momentum in the Geneva code
cannot be controlled as freely as the accretion of mass.
Strong oscillations appear in the evolution of $dJ/dM$ around the desired value when angular momentum transport
is significant in the external layers of the star, for instance when these layers are convective.
Here we compare three different values of the specific angular momentum, $\jac=1$, 3, $4\times10^{18}$ \jcgs\
with a fixed mass accretion rate of $10^{-3}$ \Mpy.
The initial conditions are given by Eq.~(\ref{eq-ini2}) and $\Omega_{\rm ini}=1$, 3, $6\times10^{-6}\rm\,[s^{-1}]$, respectively.
The results are shown in Fig.~\ref{fig-dj}.
The upper panel shows the stellar structure at such a mass accretion rate in the non-rotating case,
since rotation does not  significantly modify the stellar structure in these cases.

The oscillations in \jac\ produce oscillations in \rapv.
However, \rapv\ remains clearly in the same order of magnitude throughout the evolution.
In fact, \rapv\ shows a slight global decrease during the PMS ($M<30\,\Ms$ at this rate);
in all three cases, the initial value of \rapv\ is never exceeded during the computation.
In particular, despite an initial velocity as high as 75\% of the critical value, the fastest rotator never approaches the critical limit during its PMS evolution.
Thus in this case, no limitation in mass is encountered, and accretion can proceed freely until the zero-age main sequence (ZAMS),
without fearing the centrifugal force.

\begin{figure}\begin{center}
\includegraphics[width=0.45\textwidth]{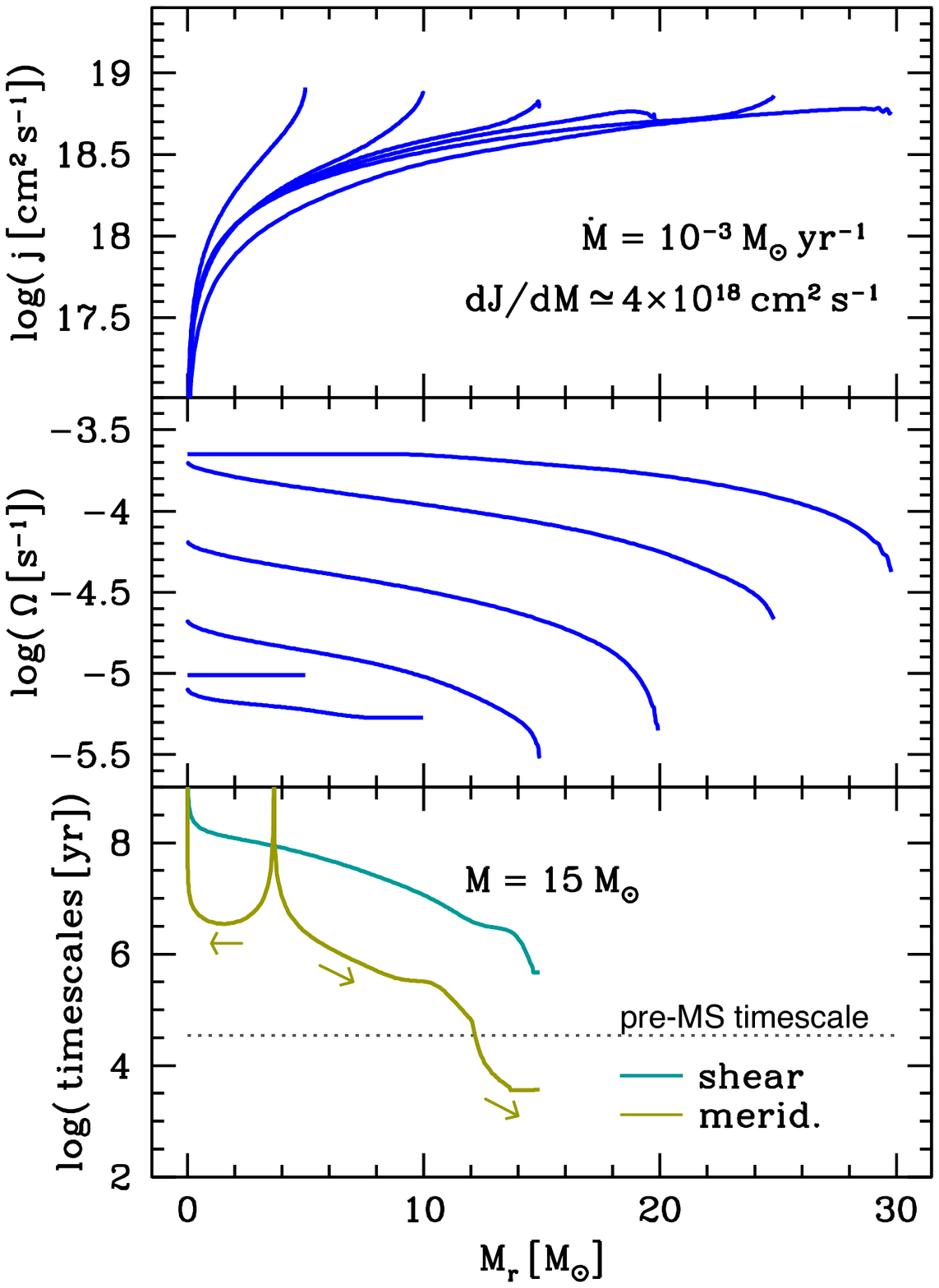}
\caption{Internal rotation profiles of the model with \dm\ = $10^{-3}$ \Mpy\ and \jac\ $\simeq4\times10^{18}$ \jcgs\ 
(fastest rotator of Fig.~\ref{fig-dj}) for $M=5$, 10, 15, 20, 25, 30 \Ms, as a function of the Lagrangian coordinate $M_r$.
The upper panel shows the profiles of the specific angular moment $j={2\over3}r^2\Omega$,
and the second one shows the profiles of the angular velocity $\Omega$.
On the bottom panel, we display the timescales for angular momentum transport by shear diffusion and meridional circulation
at $M=15$ \Ms, and compare it with the PMS timescale \Mz/\dm\ $\simeq$ 35\,000 yr.
The arrows indicate the direction of the angular momentum transport by meridional currents, outwards or inwards.}
\label{fig-djprof}
\end{center}\end{figure}

The internal rotation profiles of the fastest rotator considered here are shown in Fig.~\ref{fig-djprof}.
As in the previous cases, the more efficient process for angular momentum transport is convection
(compare the $j$-profiles at $M=5$ and 10 \Ms, and those at $M=25$ and 30 \Ms).
But in this case, some angular momentum transport also acts in radiative regions.
Indeed, by comparing the profiles at $M=15$, 20, and 25 \Ms, when the star is fully radiative (see the upper panel of Fig.~\ref{fig-dj}),
we see that $j$ is not constant for fixed $M_r$.
For instance, the value of $j$ corresponding to the layer with $M_r=15$ \Ms\ decreases between $M=15$ and 25 \Ms.
This is an effect of meridional circulation, that transports angular momentum from the contracting core outwards into the external regions.
This can be seen i the lower panel of Fig.~\ref{fig-djprof}, where we compare the timescales for angular momentum transport
by shear diffusion and by meridional currents with the timescale of the PMS phase, here given by the accretion time
\Mz/\dm\ $\simeq$ 35 000 years, where we used \Mz\ $=$ 35 \Ms\ (see the upper panel of Fig.~\ref{fig-dj}).
The profiles are shown only for $M=15$ \Ms, when the star is fully radiative.
The timescale for shear diffusion, given by $R^2/D(r)$, always exceeds the PMS time, by at least one order of magnitude,
which means that this process is negligible on such evolutionary timescales.
On the other hand, the timescale for meridional circulation, that is, the Eddington-Sweet time $R/U(r)$,
becomes shorter than the PMS time in the external layers ($M_r>12$ \Ms).
In this region, $U(r)<0$, which means that angular momentum is transported outwards.
This outwards transport is responsible for the oscillations in $dJ/dM$ (second panel of Fig.~\ref{fig-dj}).
Nevertheless, this feature only concerns the surface layers, and only becomes significant for the fast rotator considered here.
We note that $U(r)>0$ in the inner regions ($M_r<4$ \Ms), and there angular momentum is transported inwards.
However, this mechanism remains inefficient here, with a characteristic timescale two orders of magnitude longer than the PMS time.
Thus, the evolution of the rotational properties in radiative regions remains dominated by local angular momentum conservation,
and, globally, the $j$-profiles reflect the J-accretion history, as in the cases considered previously.
The present J-accretion history leads to $\Omega$-profiles that decrease outwards,
except in convective regions where it is flat (second panel of Fig.~\ref{fig-djprof}).
In contrast to the smooth-J case, the assumption of nearly constant \jac\ compensates for the spinning up of the fast contracting layers.
Indeed, these layers are accreted at large $r$, and thus a constant \jac\ implies a low $\Omega$, according to $\Omega\propto r^{-2}$.

Notice that, even for the fastest rotator considered here, rotation does not impact significantly the stellar structure or the evolutionary track on the HR diagram.
In order to estimate the deformation (flattening) of the star due to rotation,
we plotted the ratio of equatorial to polar radius on the lower panel of Fig.~\ref{fig-dj}.
The deformation only becomes significant ($R_{\rm eq}-R_{\rm pol}>10\%\,R_{\rm pol}$) for fast rotators ($\vsurf>50\%$ \vcrit),
according to the Roche model.
In the Roche approximation, the deformation of the star is modelled by including the centrifugal force in the hydrostatic equilibrium,
but without any change in the gravitational force due to the modifications in the mass distribution induced by the deformation.
This approximation gives results that are in good agreement with observed fast rotators \citep{peterson2006,maeder2009}.

\subsection{Impact of initial conditions}
\label{sec-mod-ini}

In \papI,\ we showed that the initial stellar structure significantly impacts the PMS evolution with accretion at high rates.
We found that compact radiative  (RC) initial models with a low entropy content lead to a much stronger swelling
than the extended convective (CV) initial models  of high entropy content used above.
With the RC models, the stellar radius exceeds 100~\Rs\ during the swelling, similar to the models of \cite{hosokawa2010}.
In the present section, we describe how this effect impacts the rotational history of the star.

\begin{figure}\begin{center}
\includegraphics[width=0.45\textwidth]{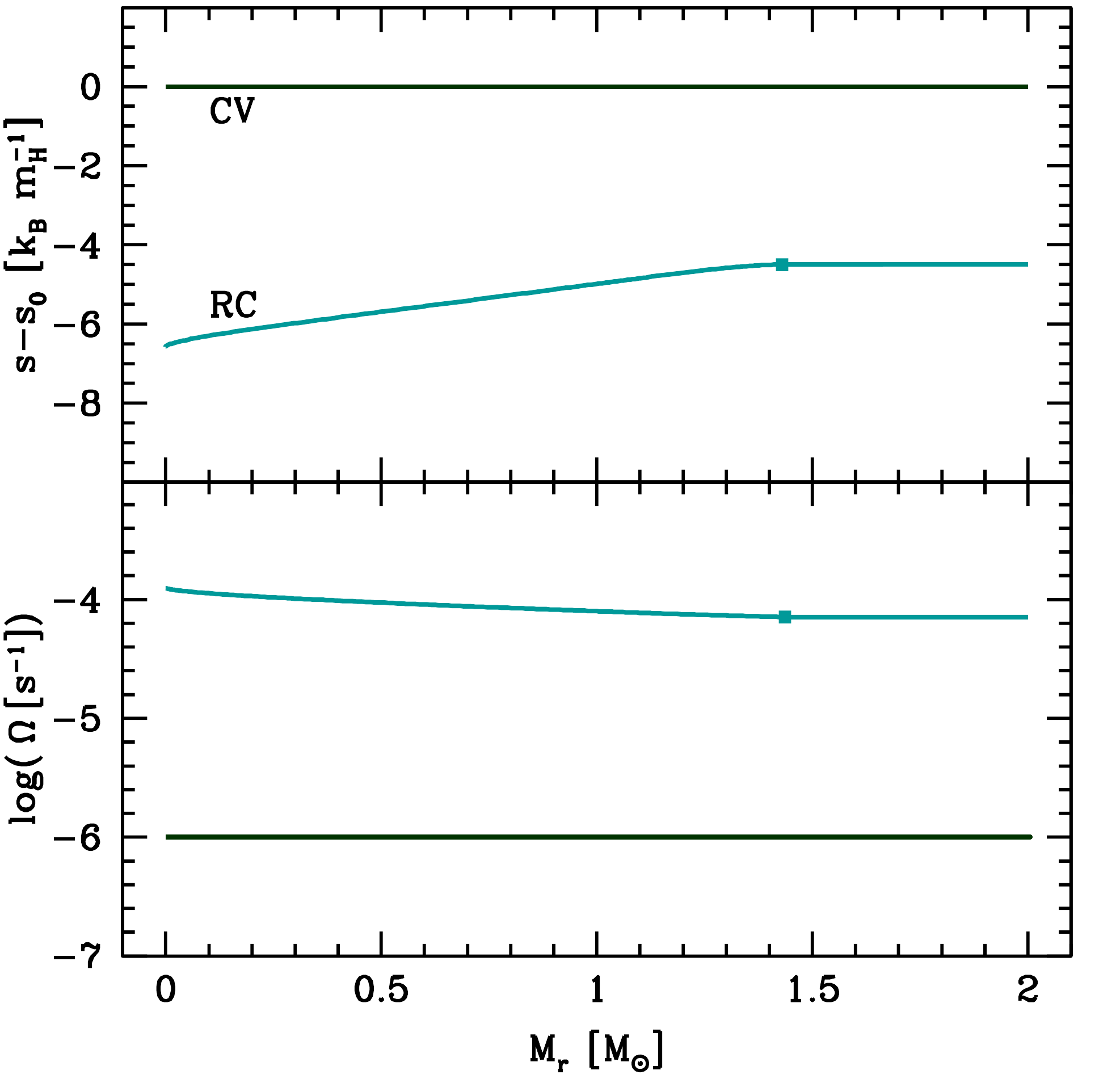}
\caption{Internal structure of the initial models described in Sect.~\ref{sec-mod-ini}.
The upper panel shows the entropy profile and the lower panel the $\Omega$-profile.
Entropy is computed from an arbitrary reference value $s_0$ that corresponds to the high entropy CV model.
In each panel, the boundary between convection and radiation in the low-entropy RC model is indicated by a dot.}
\label{fig-iniprof}
\end{center}\end{figure}

The surface properties of the CV model are given by parameters (\ref{eq-ini2}), and we consider $\Omega_{\rm ini}=10^{-6}$ s$^{-1}$,
which corresponds to the slowest rotator described in Sect.~\ref{sec-mod-dj}, with \rapv\,=\,0.13.
Our RC initial model is given by
\begin{equation}
M=2\,\Ms                \qquad  L=4.32\,\Ls\    \qquad  \Teff=5230\rm\,K
\label{eq-inir},\end{equation}
and has a radius of 2.53 \Rs.
Its structure is composed of a radiative core ($M_r<1.53\,\Ms$) surrounded by a convective envelope.
Thus, in this case, the assumption of solid-body rotation does not apply for the whole star.
We give to the RC model the same total amount of angular momentum as that of the CV model,
and the rotation profile is obtained by a constant-mass contraction from the CV model.
The entropy profiles and the $\Omega$-profiles of these two models are shown in Fig.~\ref{fig-iniprof}.
Since the RC model is much more compact than the CV model, the fact that we choose both to have the same amount of angular momentum
leads to a faster rotation in RC than in CV.
Indeed, the RC model corresponds to an initial ratio \rapv\,=\,0.4 instead of 0.1 for CV.
Moreover, we see that the $\Omega$-profile of the RC model is not flat, but shows differential rotation in the radiative core.

\begin{figure}\begin{center}
\includegraphics[width=0.45\textwidth]{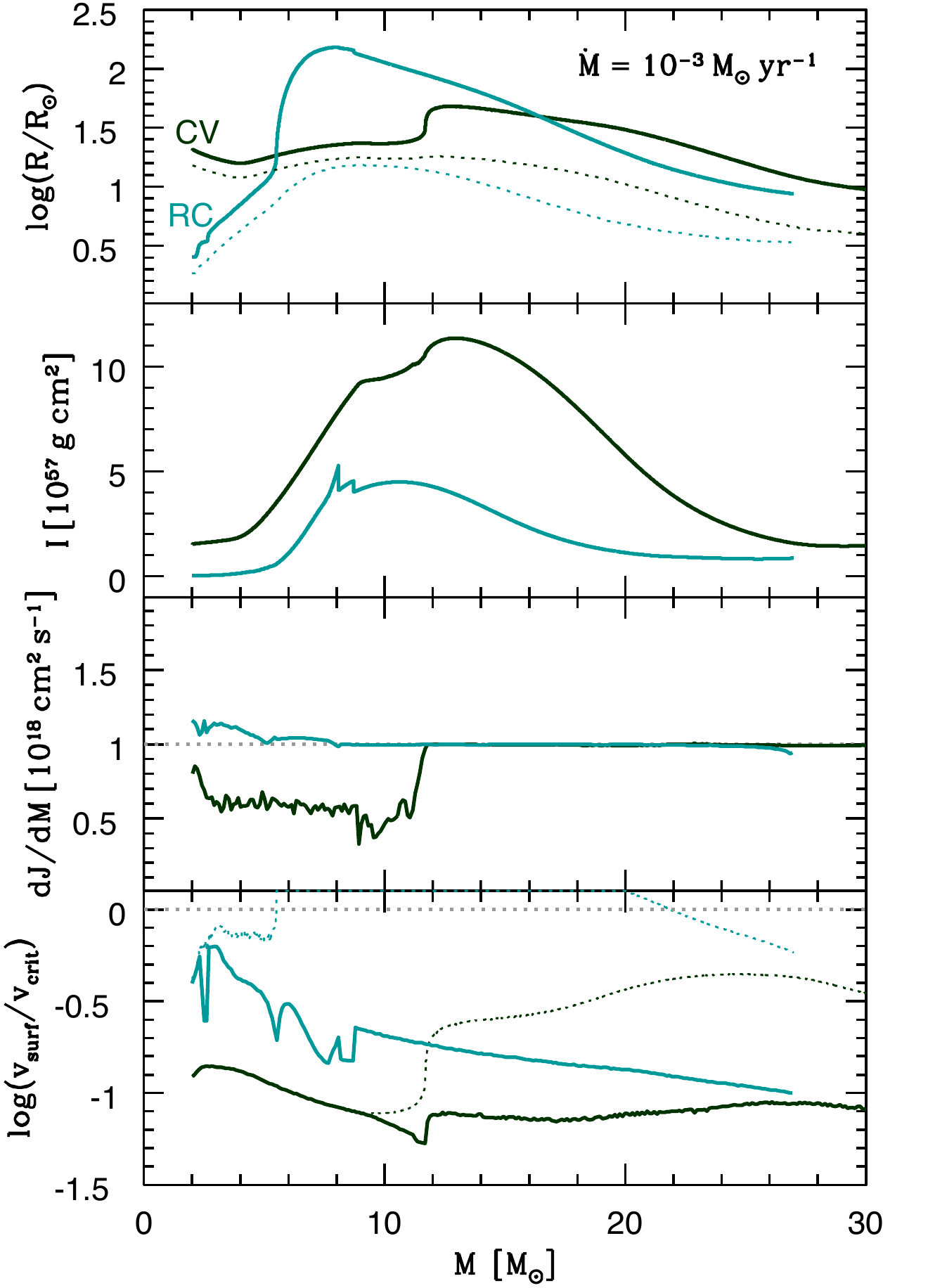}
\caption{Evolution of the models described in Sect.~\ref{sec-mod-ini},
with $\dm=10^{-3}$ \Mpy\ and $\jac\simeq10^{18}$ cm$^2$ s$^{-1}$, started from the CV and the RC models (black and green lines, respectively).
The upper panel shows the stellar radius, the second panel the moment of inertia, the third panel the accretion of angular momentum,
and the lowest panel the ratio of the surface rotation velocity to the critical velocity.
In the upper panel, the solid lines correspond to the stellar radius
and the dotted lines indicate the location of the layer that encloses 90 \% of the actual stellar mass.
In the third panel, the grey dotted line is the ideal value of $\jac=10^{18}$ cm$^2$ s$^{-1}$ while in the lowest panel, the grey dotted line indicate the critical limit
and the coloured dotted lines indicate the evolution of \rapv\ with the assumption of solid-body rotation (see Sect.~\ref{sec-bla-sam}).}
\label{fig-ini}
\end{center}\end{figure}

The evolution of the stellar radius, the moment of inertia, the accretion of angular momentum and the ratio of the surface velocity to the critical one
are shown in Fig.~\ref{fig-ini} for models computed from CV and RD with $\dm=10^{-3}$ \Mpy\ and $\jac\simeq10^{18}$ cm$^2$ s$^{-1}$.
While the radius of the CV model remains always below 50 \Rs, that of the RC model exceeds 100 \Rs\ during the swelling,
as described in \papI\ in the non-rotating case.
But, counter-intuitively, despite its large radius during the swelling, the RC model retains a much smaller moment of inertia than the CV model.
Indeed, the initial compactness of the RC model gives it a small moment of inertia.
Then, during the swelling, contraction is highly non-homologous and the mass distribution is strongly centralised;
only a small fraction of the stellar mass is contained in the extended outer layers.
Most of the mass remains close to the centre, with a small moment of inertia.
This is illustrated by the dotted lines on the upper panel, that show the radius enclosing 90 \% of the actual stellar mass;
this radius remains constantly smaller for the RC model than for the CV model, even during the swelling.

As mentioned in Sect.~\ref{sec-mod-dj}, the constant-\jac\ assumption
compensates for the variations in rotation velocity due to the swelling and the rapid contraction,
because the larger the radius, the lower the velocity of the material accreted at a given \jac.
We see in the lower panel of Fig.~\ref{fig-ini} that the swelling does not significantly affect the ratio \rapv;
despite a few oscillations, \rapv\ remains in the same order of magnitude.
In fact, the values of \rapv\ for the two models converge slowly as the star contracts towards the ZAMS.
Despite a relatively high initial velocity ($\rapv\simeq40\%$), the RC models never reach the critical limit during the PMS,
and approach the ZAMS at $\rapv\simeq10\%$, as in the CV case.
In the constant-\jac\ case, the RC model leads to a slightly stronger decrease in \rapv\ during the accretion phase,
compared to the CV model.
This fact does not result from the magnitude of the swelling,
but simply from the fact that the initial moment of inertia is smaller in the RC case than in the CV case.
Thus, producing massive stars rotating at a given surface velocity using a given angular momentum accretion history
requires a higher initial ratio \rapv\ in the RC case than in the CV case.
This fact makes it harder to produce massive fast rotators starting from RC than starting from CV,
since the initial ratio \rapv\ must always be smaller than 1.

Notice that the angular momentum transport by shear diffusion and meridional circulation remains completely negligible in the RC case
due to the decrease in \rapv\ that leads to slow rotation.
Thus, the variations in \jac\ around the specified value are lower in the RC case than in the CV case,
because the RC model is essentially radiative, with only a small convective envelope that disappears at $M=5.5$ \Ms.

\section{Birthline, PMS contraction, and MS evolution}
\label{sec-ms}

Once a star reaches its final mass and stops accreting, its evolution proceeds according to the constant mass scenario,
or with a mass that decreases with time in case of significant mass loss.
For low- and intermediate-mass stars, accretion is expected to stop before the ZAMS (\papI),
and the further PMS phase corresponds to a classical KH contraction towards the ZAMS.
In the low-mass case, the KH time exceeds the accretion time,
and the timescale of the constant mass phase is expected to be much longer than that of the accretion phase.
For massive stars, the ZAMS is expected to be reached before the end of accretion (\papI).
Thus, in this case, the MS phase directly follows the accretion phase.
In both cases, the rapid accretion phase is followed by a much slower evolution.
In the present section, we describe a birthline computed with the effect of rotation
and study how it impacts the rotational properties of stars during these slow KH or MS phases.
Observationally, the \textquoteleft birthline\textquoteright\ is defined as the upper envelope of pre-MS stars on the HR diagram.
In the context of a unique accretion history, it corresponds also to the evolutionary track of accreting stars. See \papI\ for more details.

\subsection{Birthline}
\label{sec-ms-bl}

\begin{figure}\begin{center}
\includegraphics[width=0.45\textwidth]{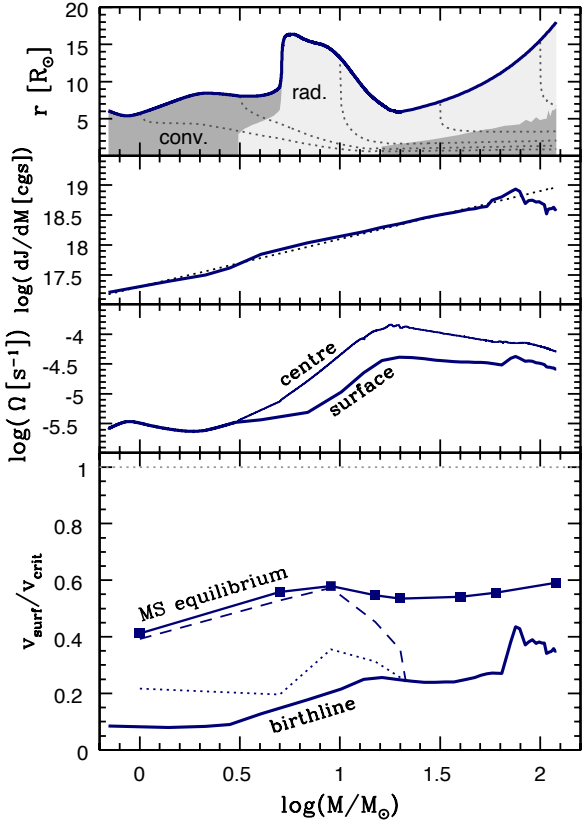}
\caption{Properties of the birthline described in Sect.~\ref{sec-ms-bl}.
The upper panel shows the stellar radius (upper blue curve) and the internal structure,
with convective regions in dark grey, radiative regions in light grey, and iso-mass as dotted curves.
The second panel shows the accretion of specific angular momentum (in logarithm of cm$^2$ s$^{-1}$):
the black dotted line is the value given by Eq.~(\ref{eq-jbl}) and the solid blue line indicates the angular momentum effectively accreted.
The third panel shows the angular velocity at the surface and in the centre.
The lower panel shows the ratio \rapv\ (lower thick solid blue line).
The filled squares connected by the thin blue line indicate the values of \rapv\ reached by the models at constant mass
computed from the birthline once they have converged to their MS equilibrium (see Sect.~\ref{sec-ms-mcst}).
The blue dashed line shows the values of \rapv\ of the same models when they reach their minimum radius corresponding to the ZAMS.
We also added the value of \rapv\ these models would have at the minimum radius if angular momentum was locally conserved (blue dotted line).
The grey dotted horizontal line indicates the critical limit.}
\label{fig-bl}
\end{center}\end{figure}

We build by accretion models for the mass range from 1 to 120 \Ms.
For simplicity, we consider only one single accretion scenario for mass and angular momentum.
As shown in \papI, the birthline given by the CH accretion law provides the best fit to the observations of T Tauri and Herbig Ae/Be stars.
Thus we choose this law for the accretion of mass.
For the accretion of angular momentum, we aim at reaching a unique value of $\rapv\simeq0.5$ on the ZAMS,
which is typical for MS stars \citep{huang2010}.
We consider an accretion history given approximately by
\begin{equation}
\jac=\left({M\over M_\odot}\right)^{0.8}\,0.2\times10^{18}\,\jcgs.
\label{eq-jbl}\end{equation}
Notice that the only justification of our choice of a unique accretion history for the angular momentum
is the fact that the prescription of Eq.~(\ref{eq-jbl}) allows us to reach $\rapv\simeq0.5$ on the ZAMS
for all the masses considered, as shown below (Fig.~\ref{fig-bl}).
In particular, the exponent 0.8 in Eq.~(\ref{eq-jbl}) is only an ad-hoc value chosen to that end.

We start the computation at 0.7 \Ms\ with initial conditions given by Eq.~(\ref{eq-mod1-ini}),
corresponding to the CV case of Sect.~\ref{sec-mod-ini}, and we stop when the stellar mass reaches 120 \Ms.
As mentioned in Sect.~\ref{sec-mod-ini}, for a constant mass accretion rate of $10^{-3}$ \Mpy,
the choice of the initial model has a significant impact on the structure of PMS stars, and then on their rotational properties.
However, in the case of CH accretion, this effect remains negligible,
since the accretion rate is still of the order of $10^{-4}$ \Mpy\ when the swelling occurs, as mentioned in \papI.
Thus we do not expect the use of RC initial conditions, instead of the CV model considered here, to significantly change the results of the present section.

The properties of the birthline we obtain with these assumptions are shown in Fig.~\ref{fig-bl}.
The evolution of the stellar radius and the internal structure are similar as in the cases described in \papI\ and in Sect.~\ref{sec-mod}.
With the CH accretion law, the ZAMS is reached slightly above 20 \Ms.
The accreted angular momentum follows the law given by Eq.~(\ref{eq-jbl}) relatively regularly until 50 \Ms.
Then, strong oscillations appear leading to a net decrease in \jac\ for masses $\sim100$ \Ms.

At the beginning of the evolution, the fully convective structure imposes solid-body rotation ($\rm\Omega_{centre}=\Omega_{surface}$),
but once the radiative core appears and the swelling starts, differential rotation develops due to the highly non-homologous contraction
(third panel of Fig.~\ref{fig-bl});
the rapidly contracting core rotates faster than the surface, which contracts at a slower rate due to the entropy it receives from the core.
At the end of the swelling, the ratio $\rm\Omega_{centre}/\Omega_{surface}$ reaches a maximum of 7.
During the KH phase that follows the swelling, the external layers can contract rapidly, and they spin up consequently.
On the ZAMS ($\log M/\Ms\simeq1.3$), however, the convective core rotates still three times faster than the surface.
After that, the expansion of the central layers due to H-burning slows down the rotation of the core,
while $\rm\Omega_{surf}$ remains almost constant (\jac\ and $R$ are both increasing).
Differential rotation is still present at the end of our computation with a core rotating two times faster than the surface.
The critical limit is never reached during the evolution (lower panel of Fig.~\ref{fig-bl}).
Despite a slight increase in \rapv\ during the accretion phase, the surface velocity remains under 50 \% of the critical value until the end of the computation.
This behaviour simply reflects the J-accretion history, since angular momentum transport by meridional circulation and shear turbulence
also remains negligible in this model.

\subsection{Evolution at constant mass}
\label{sec-ms-mcst}

From the birthline described in Sect.~\ref{sec-ms-bl}, we compute the following evolution of models of various masses by switching off accretion.
We include mass loss in these models, with the same treatment as in \cite{ekstroem2012}.
The models of various masses are thus labelled by their mass at the end of the accretion phase,
written here \Mz, since mass losses are negligible during the PMS phase.
We consider \Mz=1, 5, 9, 15, 20, 40, 60 and 120 \Ms.
Notice, however, that stars with masses above 20 \Ms\ reach the ZAMS before the end of their accretion phase (Sect.~\ref{sec-ms-bl}).
Thus, strictly speaking, \Mz\ is not the ZAMS mass for those models.

Only models with $\Mz\leq20$ \Ms\ experience a PMS contraction,
while for models with $\Mz>20$ \Ms\ , the accretion phase overlaps with early MS evolution.
In all cases, as mentioned at the beginning of the present section, the rapid accretion phase is followed by a slower evolution,
during which the angular momentum transport in radiative regions can become significant.
Indeed, it is well known that meridional currents lead to strong angular momentum redistribution during the MS;
the high efficiency of this mechanism makes the internal rotation profile converge towards an equilibrium profile
after only a few percent of the MS time \citep{zahn1992,meynet2000,maeder2001}.
While the rotation profiles at the end of the accretion phase essentially reflect the J-accretion history (see Sect.~\ref{sec-mod}),
the MS profiles are expected to be entirely rebuilt by the actions of meridional currents.
In fact, the MS equilibrium profile and the corresponding surface velocities of a model of given mass and metallicity
depend on the accretion history only through the total amount of angular momentum accreted \citep{haemmerle2013,granada2014}.

In order to show how this angular momentum redistribution impacts our models, we plotted on the lower panel of Fig.~\ref{fig-bl}
the ratio \rapv\ reached by models with various \Mz\ once they had converged to their MS equilibrium.
We see that the stellar surface of all our models spins up between the birthline and the MS equilibrium, with a gain of 0.2 -- 0.3 in \rapv.
The equilibrium values of \rapv\ for the various \Mz\ are all included in the range 40 -- 60\%,
due to our choice of the J-accretion history (see Sect.~\ref{sec-ms-bl}).
For $\Mz>20$ \Ms, since the minimum radius is reached on the birthline,
this spinning up is entirely due to angular momentum redistribution by meridional currents.
For models at lower masses, the PMS contraction between the birthline and the ZAMS  also contributes to this increase
by local angular momentum conservation.
In order to disentangle these two mechanisms, we added two additional curves on the lower panel of Fig.~\ref{fig-bl}:
the blue dashed line indicates the value of \rapv\ of the same models at the end of the PMS contraction (minimum radius),
and the blue dotted line shows the same without the effect of angular momentum transport,
computed simply by using $\vsurf\times R_{\rm min}={\rm v_{birthline}}\times R_{\rm birthline}$.
In other words, the dashed line delimitates the PMS and the MS spinning up,
while the dotted line delimitates the effect of local angular momentum conservation and of angular momentum transport.
As we see, in the whole mass range, the spinning up is dominated by the angular moment transport from meridional currents.
Moreover, in the low- and intermediate-mass range, meridional currents are efficient enough
to achieve convergence towards the MS equilibrium profiles already during the PMS contraction.
Thus, during the PMS contraction and the MS evolution, the internal angular momentum transport regulates the rotational history of the star,
in contrast to the accretion phase.

\begin{figure}\begin{center}
\includegraphics[width=0.45\textwidth]{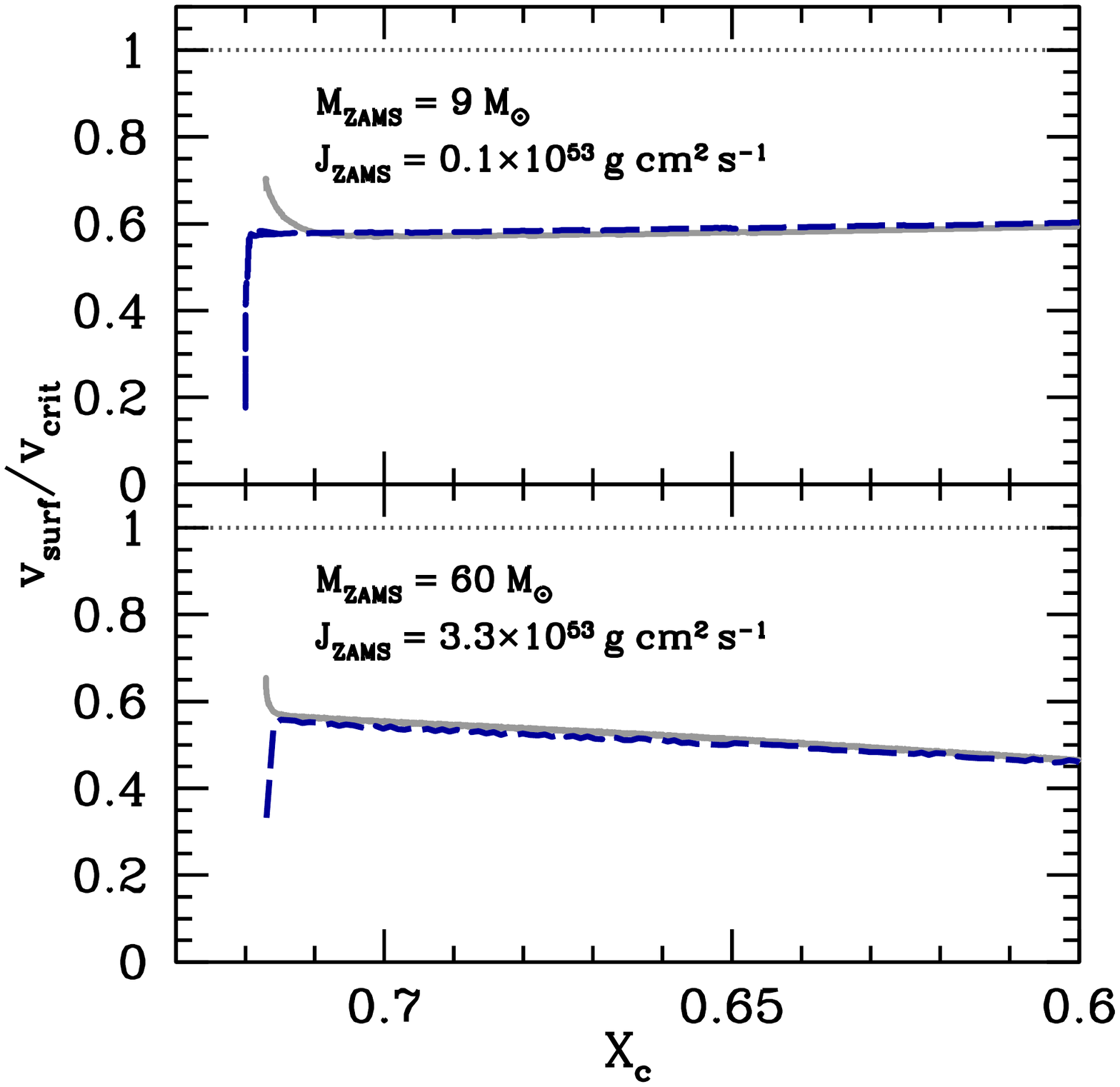}
\caption{Convergence of the surface velocities towards MS equilibrium.
Both panels show the evolution of the ratio \rapv\ during the early MS as a function of the central mass-fraction of hydrogen that depicts time along the MS.
The upper panel shows the case \Mz\ = 9 \Ms\ and the lower panel the case \Mz\ = 60 \Ms.
On each panel, the blue dashed curve corresponds to the model built by accretion, described in Sect.~\ref{sec-ms-mcst},
while the grey solid curve indicates a model started on the ZAMS with a flat rotation profile
with the same mass and angular momentum as the blue model.
The grey dotted horizontal lines indicate the critical limit.}
\label{fig-mcst}
\end{center}\end{figure}

Finally, in order to illustrate the convergence process of rotation profiles during the early MS,
we computed models starting from the ZAMS with flat rotation profiles, with the same total amount of angular momentum as our models built by accretion.
Fig.~\ref{fig-mcst} shows the cases where \Mz\ = 9 and 60 \Ms.
Despite the differences in the initial surface velocities, we see that models with the same mass and angular momentum
converge rapidly towards the same value of \rapv.
In the case of models built by accretion, and with highly differential rotation profiles on the ZAMS,
meridional currents redistribute the angular momentum from the fast rotating core to the slowly rotating surface, leading to an increase in \rapv.
In contrast, for models starting on the ZAMS with solid-body rotation, meridional currents redistribute the angular momentum from the surface to the core.
Due to the short timescale of the convergence process, the differences in the rotation profiles do not survive long enough
to have a significant impact on further evolution in terms of chemical composition or evolutionary tracks.
This result was already described in detail in \cite{haemmerle2013}.
Since it is based on the actions of meridional currents after the accretion phase,
the improvements brought to the code since this work have no impact on this conclusion.

\section{Discussion}
\label{sec-bla}

\subsection{The angular momentum barrier}
\label{sec-bla-lim}

Stellar models described in Sect.~\ref{sec-mod} show that smooth-J accretion leads to a limitation in the stellar mass;
if the accreted material is dragged by the stellar surface without any braking, the critical rotation limit is reached before the star becomes massive.
This {\it angular momentum barrier} is independent of the choice of the initial rotation velocity (Sect.~\ref{sec-mod-vini})
or of the mass accretion history (Sect.~\ref{sec-mod-dm}).
Only a change in the J-accretion history can compensate this effect and overcome any limitation in mass or velocity during the accretion phase.
We showed in Sect.~\ref{sec-mod-dj} that the constant-\jac\ assumption with $\jac\simeq1-4\times10^{18}$ \jcgs\
allows for circumvention of the angular momentum barrier.
However, the physical conditions that lead to such angular momentum accretion rates remain unknown.
How the accreted angular momentum compares with the Keplerian value
\begin{equation}
\jk=\sqrt{GMR}
\label{eq-jk}\end{equation}
expected in the accretion disc also remains to be explored.

From the mass-radius relation of the birthline described in Sect.~\ref{sec-ms-bl} (upper panel of Fig.~\ref{fig-bl}),
one can estimate the typical mass and radius of PMS stars accreting at high rates as 10 \Ms\ and 10 \Rs, respectively.
This leads to
\begin{equation}
\jk\sim\sqrt{G\times10\,\Ms\times10\,\Rs}\simeq3\times10^{19}\,\jcgs.
\label{eq-jknum}\end{equation}
Comparing this value to the angular momentum effectively accreted (second panel of Fig.~\ref{fig-bl}),
we see that \jac\ remains lower than 10\% of the Keplerian angular momentum.
The same conclusion holds for the models described in Sect.~\ref{sec-mod-dj}.
In Appendix~\ref{app-kep}, we show that, indeed, the angular momentum accreted has to be lower than one third of the Keplerian angular momentum.
This indicates that the angular momentum barrier can only be circumvented with a strong braking mechanism,
efficient enough to remove more than two thirds of the angular momentum from the Keplerian disc.

\subsection{The key-role of the J-accretion law}
\label{sec-bla-dj}

Models of Sect.~\ref{sec-mod} also show that angular momentum transport by shear diffusion and meridional circulation
remains weak in the short timescale of the accretion phase,
and that the evolution of the rotational properties of accreting stars is dominated by local angular momentum conservation.
As a consequence, nearly any rotation profile can be built during the accretion phase by choosing the appropriate J-accretion history.
Indeed, each Lagrangian layer retains essentially the same angular momentum it had when it was accreted,
so that the $j$-profiles reflect the J-accretion history.
Then, the $\Omega$-profiles are given by the $j$-profiles and the relative contraction of each layer,
which is almost independent of rotation and can be taken from non-rotating models.
Angular momentum transport by meridional currents can only modify this fact
in the very external layers of fast rotators (Sect.~\ref{sec-mod-dj}).

An exception to this rule is the case of convective zones, in which angular momentum transport is instantaneous because of the solid-body assumption.
Convection appears in two regions in our models, first in the convective envelope before the swelling, then in the convective core after the ZAMS.
During the crucial phase of the post-swelling KH contraction, all the fast contracting external layers
that reach the critical limit in the smooth-J case have been radiative since they were accreted.
This explains why a change in the accretion law solves the problem of the angular momentum barrier;
if \jac\ is kept constant, this means that $\Omega$ will be small when material is accreted at large radii during the swelling, according to Eq.~(\ref{eq-omac}).
When the star contracts, the increase of $\Omega$ will then not be sufficient for $\Omega$ to reach the critical limit.

Interestingly, we notice that for stars that follow the birthline until the ZAMS, all the layers of the convective envelope
are eventually incorporated into the convective core (see the iso-mass in the upper panel of Fig.~\ref{fig-bl}).
As a consequence, all the layers of massive stars that are radiative at the end of the accretion phase have been so since they were accreted,
and thus they experienced only weak angular momentum transport.
So, except for the constraint of solid-body rotation in the convective core, the rotation profiles of massive stars at the beginning of the MS
reflect essentially the J-accretion history.

For low- and intermediate-mass stars, we showed in Sect.~\ref{sec-ms-mcst} that angular momentum transport by meridional currents
is strong enough to dominate the rotational history during the PMS contraction that follows the accretion phase.
However, the action of these meridional currents simply leads to a convergence of the rotation profiles towards MS equilibrium,
and therefore we expect this action to be null if the rotation profile at the end of the accretion phase is close enough to equilibrium.

The freedom in the early MS rotation profile of massive stars has interesting consequences for the evolution of rotation on the MS,
in particular for fast rotators and Be stars.
\cite{granada2014} presented a grid of MS and post-MS models that rotate close to their critical limit during the whole MS.
Such models were lacking in previous grids due to the usual assumption of solid-body rotation on the ZAMS.
Indeed, as described in Sect.~\ref{sec-ms-mcst}, with a flat rotation profile on the ZAMS,
meridional currents lead to a decrease in the surface velocity during the early MS.
Thus even models started on the ZAMS at the critical velocity evolve rapidly towards $\vsurf<80\%\,\vcrit$ \citep{ekstroem2008a}.
In order to produce models that account for the fast rotators observed throughout the MS \citep{huang2010},
one needs to consider differential rotation upon arrival on the ZAMS.
Indeed, the justification of solid-body rotation on the ZAMS is based on the assumption that PMS stars are convective.
However, it is well known that large radiative regions develop already during the PMS
(\citealt{stahler1980a,stahler1980b,hosokawa2009,hosokawa2010}; see also \papI),
and our computations demonstrate that non-homologous contraction of the various layers leads to significant differential rotation before the star reaches the ZAMS.
The non-flat rotation profiles used by \cite{granada2014} on the ZAMS came from two sources;
some were taken from PMS computations \citep{haemmerle2013}, performed with the old version of the Geneva code,
that were slightly differentially rotating.
However, differential rotation in these models was still too low compared to the MS equilibrium profiles,
and ad-hoc profiles were needed to account for the fastest rotators, having $\vsurf\simeq95\%\,\vcrit$ during the whole MS.
The results of the present work support the validity of these two types of profiles,
and we propose that there exists an angular momentum accretion history that produces such rotation profiles on the ZAMS.
We postpone the exact determination of such an angular momentum accretion history to a later study.

\subsection{Constraints on PMS evolution from rotation}
\label{sec-bla-ini}

We showed in Sect.~\ref{sec-mod-ini} that in the constant-\jac\ case, the decrease in \rapv\ during the accretion phase
is more pronounced when using low-entropy RC initial models than high-entropy CV models.
With the RC initial model, starting at a large fraction of the critical velocity, we reach the high-mass range at a low fraction of the critical velocity.
A star that reaches 2 \Ms\ with a RC structure can rarely become a fast rotator once it enters the high-mass range,
because a ratio $\rapv<1$ at 2 \Ms\ would lead to $\rapv\ll1$ in the high-mass range.
This effect could be compensated by a change in the angular momentum accretion history (see Sect.~\ref{sec-bla-dj}),
but we see that the use of the RC initial model reduces the freedom in the J-accretion history
for the formation of massive fast rotators by accretion compared to the CV case.

Moreover, due to their compactness, RC initial models rotate closer to their critical limit
compared to CV initial models having the same total amount of angular momentum.
In the cases described in Sect.~\ref{sec-mod-ini}, we had to assume a slow rotator in the CV case
in order to be able to produce a RC initial model having the same angular momentum content and rotating under the critical limit.
A RC initial model having the same angular momentum content as the fastest rotators considered in Sect.~\ref{sec-mod-dj},
for instance, would exceed the critical limit.
In Appendix~\ref{app-kep}, we show that RC models accreting angular momentum according to $\jac=20\%\jk$ reach the critical limit at $M=5$ \Ms.
Thus one may wonder if such compact RC models could be formed by accretion without facing the critical limit.

Answering these questions requires following the early accretion phase with a consistent treatment of accretion and rotation.
This issue will be the topic of future work, but the present results already provide evidence in favour of our CV initial models.

\subsection{Semi-analytical interpretation}
\label{sec-bla-sam}

If one neglects any internal angular momentum transport and the deformation of the star due to rotation,
the evolution of the surface velocity is given simply by the accreted angular momentum, according to
\begin{equation}
\jac=j_{\rm surf}={2\over3}\ R\times\vsurf.
\label{eq-jsam}\end{equation}
As in Eq.~(\ref{eq-omac}), the factor 2/3 reflects the fact that the material at the surface is distributed over a spherical shell.
Using $\vcrit=\sqrt{GM/R}$ and Eq.~(\ref{eq-jk}), one has
\begin{equation}
\rapw={{3\over2}\jac\over\sqrt{GMR}}={3\over2}{\jac\over\jk}.
\label{eq-rapsam}\end{equation}
We note that the fraction 3/2 survives in the right-hand side of Eq.~(\ref{eq-rapsam}) because
the angular momentum from the disc, which we assume to be confined to the equatorial plan,
is redistributed spherically on the stellar surface when it is accreted, according to shellular rotation.
Indeed, the fraction 2/3 appears in the expression of $j_{\rm surf}$ (Eq.~\ref{eq-jsam}),
but not in the expression of \jk\ (Eq.~\ref{eq-jk}), therefore it does not cancel in Eq.~(\ref{eq-rapsam}).

Since the stellar mass increases by one or two orders of magnitude between our initial models and massive stars,
for constant \jac\ , the ratio \rapv\ decreases slightly during the evolution.
It remains, however, in the same order of magnitude, thanks to the square root in Eq.~(\ref{eq-rapsam}).
This behaviour is similar to that of models described in \cite{haemmerle2013},
computed with the old version of the code and the smooth-J assumption, where the mean specific angular momentum was also constant.
In contrast, however, with the new version of the code, fast rotators in the high-mass range can be produced as well
by choosing a J-accretion history in which \jac\ increases during the evolution, such as the one used in Sect.~\ref{sec-ms}.

If we assume solid-body rotation instead of local conservation of angular momentum, we have
\begin{equation}
J=I\ \Omega={I\over R}\ \vsurf  \quad\Longrightarrow\quad
\rapw=\sqrt{R^3\over GM}\ {J\over I}.
\end{equation}
Defining the dimensionless parameter
\begin{equation}
\alpha:={5\over2}{I\over MR^2}=\ 5\int_0^1{\rho\over\bar\rho}\,s^4ds
\qquad  \left(\ \bar\rho:={M\over{4\over3}\pi R^3}      ,\      s:={r\over R}\ \right)
,\end{equation}
that reflects the internal mass distribution, one has
\begin{equation}
\rapw={5\over2\alpha}{J/M\over\sqrt{GMR}}={5\over2\alpha}{{1\over M}\int_0^M\jac\,dM\over\jk}.
\label{eq-valp}\end{equation}
If \jac\ = cst, we obtain a similar relation to that in Eq.~(\ref{eq-rapsam}), with a factor $5/2\alpha$ in front, instead of 3/2.
The parameter $\alpha$ is 1 for homogeneous density distribution ($\nabla\rho=0$), and smaller for any realistic structure ($\nabla\rho<0$).
Thus, solid-body rotation increases the ratio \rapv\ compared to the differentially rotating case.
Indeed, angular momentum is more centralised in the differentially rotating case than for solid-body rotation,
due to the negative $\Omega$ gradient, as visible in Fig.~\ref{fig-djprof}, and it results in a slower surface rotation.

This fact is visible from the lower panel of Fig.~\ref{fig-ini}.
In addition to the evolution of the ratio \rapv\ for our differentially rotating models (solid lines),
we plotted the evolution of \rapv\ computed with the solid-body assumption, using $\Omega=J/I$ (dotted lines).
As soon as differential rotation appears, the solid-body case leads to higher surface velocities than in the differentially rotating case.
This effect is particularly catastrophic in the RC case, where the solid-body model reaches the critical limit at 5~\Ms.
Solid-body rotation could arise in the presence of a strong magnetic field,
for instance in the case of an efficient Tayler-Spruit dynamo, not considered here.
The present analysis shows that, in this case, the critical limit would be reached earlier than in the non-magnetic case,
that is, that the constraint of the angular momentum barrier would be stronger than in our models.

\subsection{Comparison with previous studies}
\label{sec-bla-lee}

\cite{lee2016} studied the evolution of rotation during the accretion phase in the case of Population III stars,
considering rates as high as $4\times10^{-3}$ \Mpy.
However, differential rotation was not included in their study, and the authors assumed solid-body rotation,
post-processed from computations of accreting stars performed with the MESA code \citep{paxton2011,paxton2013,paxton2015}.
The accreted angular momentum was taken as a constant fraction of the Keplerian angular momentum (given by Eq.~\ref{eq-jk}).
Stars accreting at the high rates considered in this work are known to approach the Eddington limit
before it has contracted to the ZAMS \citep{omukai2001,omukai2003}.
In the presence of rotation, the centrifugal force lowers the upper limit in luminosity, which is compatible with hydrostatic equilibrium.
This effect is known as the $\Omega\Gamma$-limit \citep{maeder2000}.
The models of \cite{lee2016} reached the $\Omega\Gamma$-limit during the swelling, which occurs at masses between 20 and 40 \Ms\
at this accretion rate, preventing further accretion and limiting the mass of stars that can form by accretion.

Our models at solar metallicity show that differential rotation cannot be neglected during the swelling phase.
The semi-analytic considerations of Sect.~\ref{sec-bla-sam} show that differential rotation can significantly reduce the ratio \rapv.
We showed that solid-body rotating models reach the critical limit if the swelling exceeds 100 \Rs,
but that differential rotation avoids such a catastrophic end.
It is thus necessary to take into account differential rotation in order to test the validity of such mass limitations.

\section{Conclusions}
\label{sec-outro}

After having improved the treatment of accretion in the Geneva code (Sect.~\ref{sec-code-imp} and Appendix~\ref{app-imp}; see also \papI),
we described in the present work the PMS evolution of solar metallicity stars with accretion and rotation
by considering various accretion laws for the mass and the angular momentum (Sect.~\ref{sec-mod}).
Our main results can be summarised as follows:

\begin{enumerate}

\item Angular momentum transport by shear and meridional currents inside the radiative layers of the star
remains weak during the short accretion timescale (Sect.~\ref{sec-mod-1}).
Only convective transport is efficient (solid-body assumption),
and the evolution of the rotational properties in radiative regions is dominated by local angular momentum conservation.
Since large radiative regions develop during the accretion phase,
the internal rotation profiles during this phase essentially reflect the angular momentum accretion history.
The highly non-homologous PMS contraction leads, in general, to significant differential rotation in the stellar interior.

\item Under the assumption of smooth angular momentum accretion,
the new version of the code leads to a limitation in mass instead of a limitation in rotation velocity (Sect.~\ref{sec-mod-1}).
This implies that if the accreted material is dragged by the stellar surface without braking,
the forming star reaches the critical limit before it can significantly enter into the high-mass range.
This is due to the rapid contraction of the stellar surface after the PMS swelling.
When the critical rotation limit is reached, accretion stops, thus preventing the formation of massive stars.
This angular momentum barrier is independent of the choice of the initial rotation velocity (Sect.~\ref{sec-mod-vini})
or of the mass accretion history (Sect.~\ref{sec-mod-dm}).

\item Only a change in the angular momentum accretion history can circumvent the angular momentum barrier,
and overcome any limitation in mass or velocity during the accretion phase.
A careful choice of the J-accretion history allows for production of stars of any mass and velocity compatible with the structure equations.
Using the constant-\jac\ assumption, the ratio \rapv\ remains of the same order of magnitude during the accretion phase,
similarly as with the old version of the code and smooth-J accretion (Sect.~\ref{sec-mod-dj}).
Angular momentum accretion histories compatible with massive star formation by accretion imply an orbital velocity in the inner disc
that is lower than one third of the Keplerian value (Sect.~\ref{sec-bla-lim} and Appendix~\ref{app-kep}).
We conclude that
{ massive star formation by accretion requires an efficient braking mechanism in the inner accretion disc in order to circumvent the angular momentum barrier}.

\item For angular momentum accretion histories that allow massive stars to form,
the internal rotation profiles during the accretion phase decrease outwards (Sect.~\ref{sec-mod-dj} and \ref{sec-ms-bl}).
The degree of differential rotation reaches its maximum after the swelling, with a core rotating approximately seven times faster than the surface.
Then, differential rotation is reduced as the stellar surface contracts, but survives until the end of our computations,
with a core that rotates two to three times faster than the surface as the star reaches a MS structure.
Such differential rotation allows the star to reach higher masses than in the case of solid-body rotation
that would eventually arise in the presence of a strong magnetic field (Sect.~\ref{sec-bla-sam}).

\item The compact initial conditions, with a low entropy content, that lead to a significant swelling during the accretion phase,
reduce the freedom in the angular momentum accretion histories that allow massive fast rotators 
to form (Sect.~\ref{sec-mod-ini} and \ref{sec-bla-ini}).
This result supports the validity of our extended initial models with high entropy content, that avoid a significant PMS swelling (see \papI).

\item The accretion history during the PMS phase has no significant impact on the MS and post-MS evolution (Sect.~\ref{sec-ms-mcst}).
Models based on various PMS scenarios converge to an internal rotation structure that depends only on their mass, their chemical composition,
and their total amount of angular momentum, after only a few percent of the MS timescale.

\item Thanks to the low efficiency of angular momentum transport by shear and meridional currents,
very different rotation profiles on the ZAMS are possible, strongly depending on the angular momentum accretion history the star experienced (Sect.~\ref{sec-bla-dj}).
This fact supports the validity of previously published models that rotate close to their critical limit during their whole MS phase.

\end{enumerate}

\appendix

\section{Effects of the improvements in the code}
\label{app-imp}

Here we describe the effects of the improvements in the code that are related to rotation.
We consider the same physical conditions as in Sect.~\ref{sec-mod-1}:
the initial model is given by Eq.~(\ref{eq-mod1-ini}) and (\ref{eq-mod1-vini}),
mass is accreted at $10^{-5}\,\Mpy$ (Eq.~\ref{eq-mod1-dm})
and angular momentum according to the smooth-J assumption, as in \cite{haemmerle2013}.
The model computed with these inputs and the new version of the code has been described in Sect.~\ref{sec-mod-1}.
Here we compare this model with the equivalent one computed with the old version of the code, using the same physical inputs.

\begin{figure}\begin{center}
\includegraphics[width=0.45\textwidth]{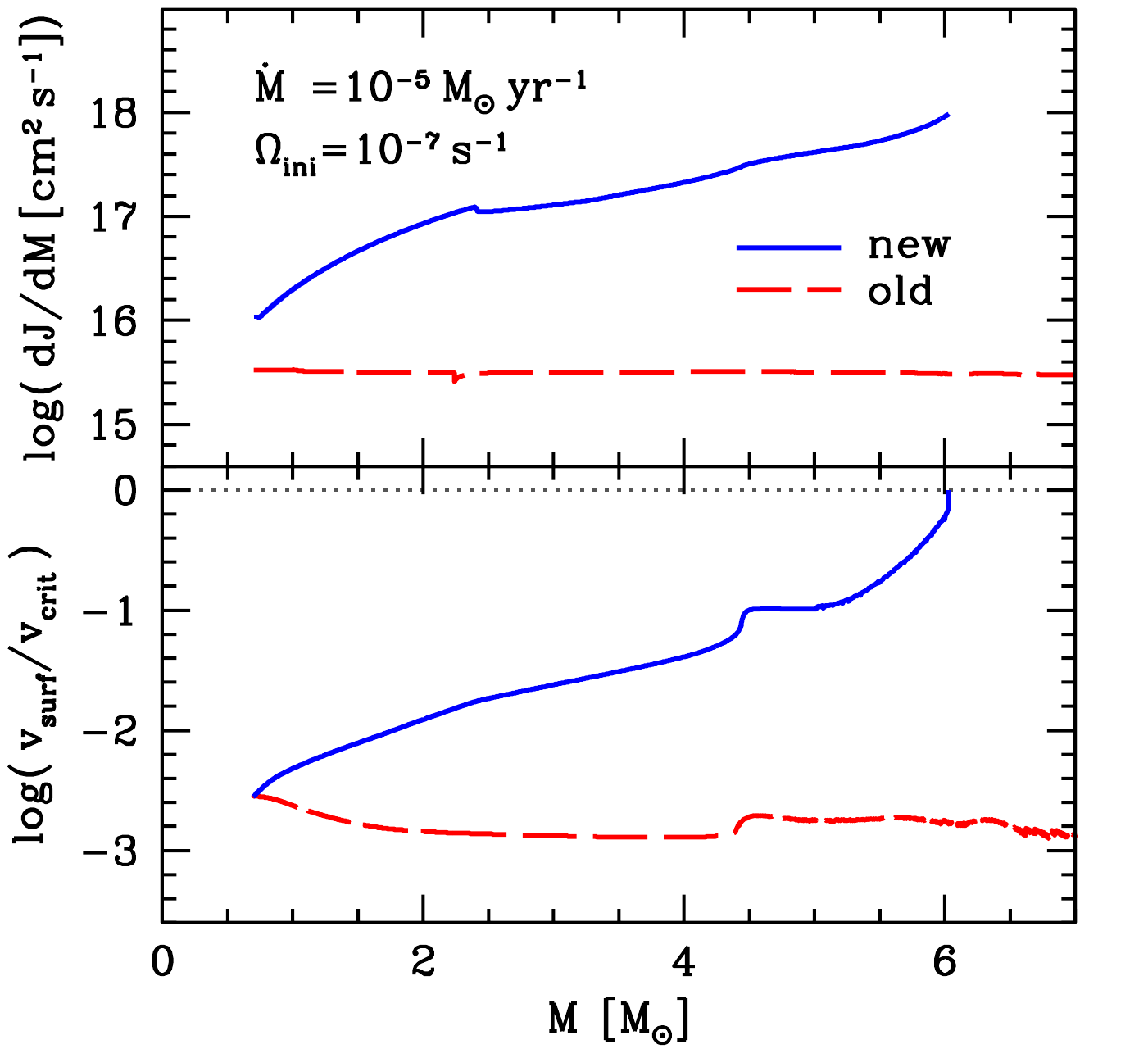}
\caption{Global rotational properties of the models described in Sect.~\ref{app-imp}.
The upper panel shows the angular momentum accretion rate and the lower panel shows the ratio \rapv,
with the critical limit indicated by the dotted horizontal grey line.}
\label{fig-imp}
\end{center}\end{figure}

The global rotational properties of these two models are compared in Fig.~\ref{fig-imp}.
Both models start their evolution at $\rm\vsurf=0.42\,km\,s^{-1}=0.28\%\,\vcrit$, according to Eq.~(\ref{eq-mod1-vini}).
In the model computed with the old version of the code, with the artificial loss of angular momentum,
the surface velocity remains in the same order of magnitude during the whole PMS,
with values in the range $\rm0.25\,km\,s^{-1}<\vsurf\lesssim1.05\,km\,s^{-1}$.
The corresponding angular momentum accretion rate is constant during the evolution ($dJ/dM\simeq3\times10^{15}$ \jcgs).
This result is in agreement with the case described in \cite{haemmerle2013}, corresponding to a different mass accretion rate
and a higher initial velocity ($\Omega\rm_{ini}=2.1\times10^{-5}\,s^{-1}$ , that is $\rapv=80\%$).

However, the behaviour of the model computed with the new version of the code, without any artificial loss of angular momentum, is drastically different.
As described in Sect.~\ref{sec-mod-1}, the surface velocity does not remain in the same order of magnitude,
but increases by several orders of magnitude, in particular during the post-swelling contraction.
The corresponding angular momentum accretion rate is no longer constant,
but increases with mass and time, by two orders of magnitude between 0.7 and 6 \Ms.
This illustrates the fact that in cases of smooth-J accretion,
the angular momentum accretion rate depends itself on the evolution of rotation in the model.

\begin{figure}\begin{center}
\includegraphics[width=0.45\textwidth]{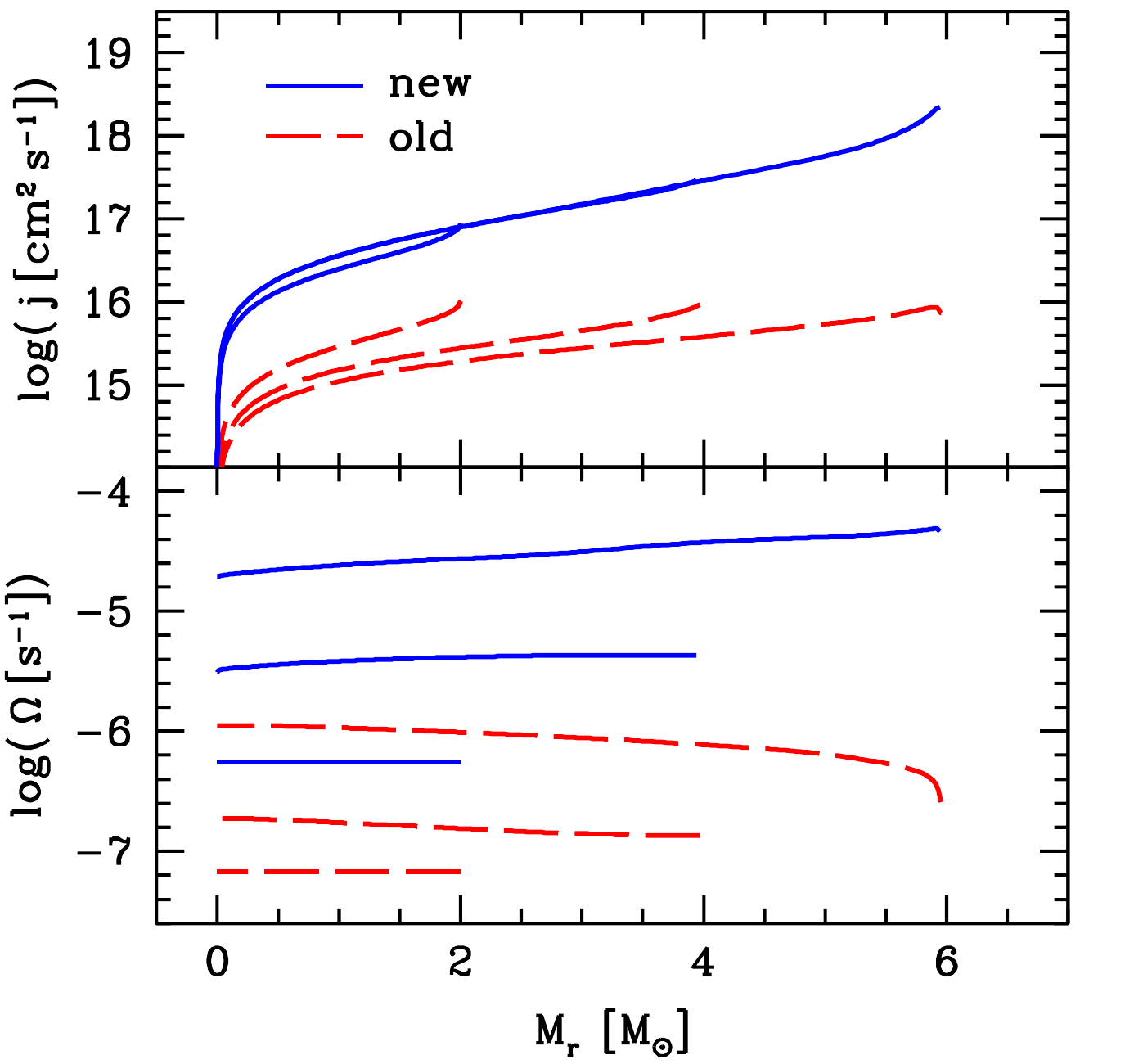}
\caption{Rotation profiles of the models described in Sect.~\ref{app-imp}, at $M=2$, 4 and 6 \Ms.
The upper panel shows the $j$-profiles and the lower panel shows the $\Omega$-profiles.}
\label{fig-improf}
\end{center}\end{figure}

Fig.~\ref{fig-improf} shows the internal rotation profiles of these two models: The $j$-profiles in the upper panel and the $\Omega$-profiles in the lower panel.
With the old version of the code, the specific angular momentum decreases with time in each layer,
that is, for each fixed value of the Lagrangian coordinate.
This is the effect of the artificial loss of angular momentum described in Sect.~\ref{sec-code}.
With the new version of the code, the angular momentum is locally conserved in each radiative layer,
which leads to $\Omega$-profiles that increase outwards, as described in Sect.~\ref{sec-mod-1}.
In the old model, the artificial loss of angular momentum
is more important in the external regions than in the core (see \citealt{haemmerle2014}).
As a consequence, the $\Omega$-profiles decrease outwards in this case.
As described in \cite{haemmerle2013}, this leads to a rotation profile on the ZAMS
that is close to the equilibrium profile with respect to meridional circulation during the MS evolution.

\section{Accretion of angular momentum at a constant fraction of the Keplerian value}
\label{app-kep}

\begin{figure}\begin{center}
\includegraphics[width=0.49\textwidth]{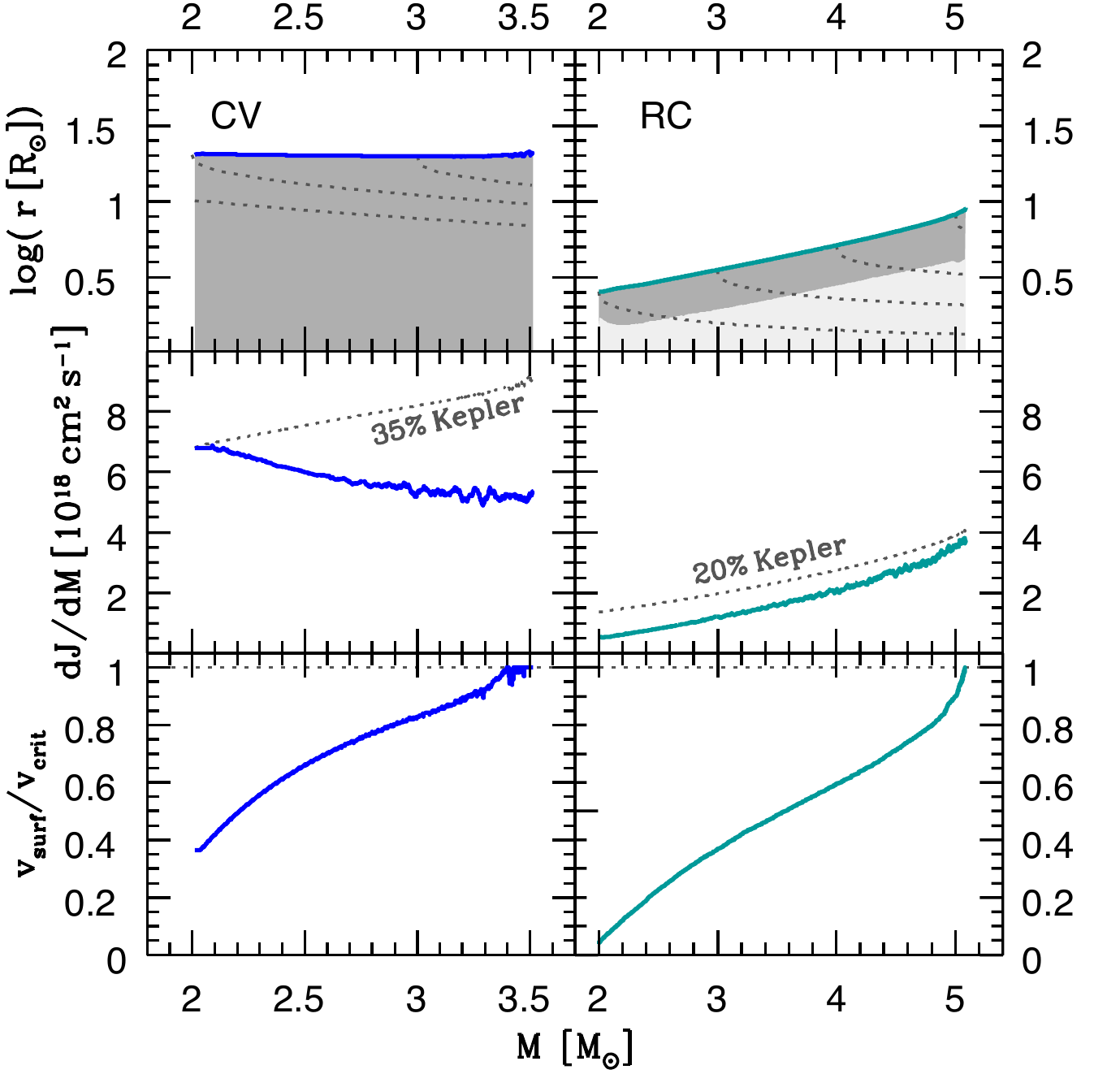}
\caption{Internal structure (upper panel), angular momentum accretion rate (middle panel), and ratio \rapv\ (lower panel)
of the models described in Sect.~\ref{app-kep}.
The left-hand column shows the properties of the models started from a CV initial structure,
while the right-hand column corresponds to the RC initial structure.
On the top panel, the upper solid curve is the stellar radius, while convective and radiative zones are shown in dark and light grey, respectively.
The dotted black curves are iso-mass at intervals of 1 \Ms.
In the middle panel, the coloured lines indicate the angular momentum effectively accreted by the models,
and the black dotted curves show the value of \jac\ corresponding to the indicated fraction of \jk.
On the lower panel, the horizontal dotted line indicates the critical limit.}
\label{fig-djk}
\end{center}\end{figure}

The models described in Sect.~\ref{sec-mod-dj}, that circumvent the angular momentum barrier,
correspond to an angular momentum accreted that is lower than 10\% of the Keplerian angular momentum \jk\ (Sect.~\ref{sec-bla-lim}).
Here we estimate the upper fraction of \jk\ ; the star can accrete without facing the angular momentum barrier.
We aim to compute models accreting angular momentum at a constant fraction of \jk.
However, as in Sect.~\ref{sec-mod-dj}, we cannot freely control the accretion of $J$ in this case;
in particular when large external regions of the star are convective.
Fig.~\ref{fig-djk} shows the stellar structure, the angular momentum accretion rate, and the ratio \rapv\
for two models accreting mass at a rate of $10^{-3}$ \Mpy.
The first one is computed from the CV initial structure, and the second one from a RC structure (see Sect.~\ref{sec-mod-ini}).

The angular momentum accreted in the CV model is always lower than 35\% of \jk.
Without any angular momentum transport, the surface velocity would thus remain approximately one third of \vcrit.
However, during the early phase, the star is fully convective and angular momentum is instantaneously redistributed in the stellar interior.
The assumption of solid-body rotation leads to outwards transport of angular momentum that produces an increase in \vsurf.
As evolution proceeds, the ratio \rapv\ increases monotonously and reaches the critical limit at a mass of 3.4 \Ms,
before the formation of the radiative core.
This example shows that the material of the accretion flow has to be slowed down to under one third of the Keplerian velocity expected in the disc.
We emphasise that this value is only an upper limit.
The exact determination of the fraction of angular momentum that must be removed from the Keplerian disc
will be the topic of a forthcoming paper.
We note, however, that according to Eq.~(\ref{eq-jknum}), the fastest rotator of Sect.~\ref{sec-mod-dj} accretes angular momentum
at $\sim10\%\jk$ without reaching the critical limit.

Since this limitation is due to the instantaneous redistribution of $J$ while the star is fully convective,
one wonders whether the RC models allow for circumvention of this limitation.
The right-hand side of Fig.~\ref{fig-djk} shows that this is not the case.
Even with $\jac<20\%\,\jk$, the angular momentum transport in the small external convective zone leads the star to the critical limit soon after 5 \Ms.

\begin{acknowledgements}
Part of this work was supported by the Swiss National Science Foundation.
LH and RSK were supported by the European Research Council under the European Community's Seventh Framework Programme (FP7/2007 - 2013)
via the ERC Advanced Grant `STARLIGHT: Formation of the First Stars' (project number 339177).
\end{acknowledgements}

\bibliographystyle{aa}
\bibliography{bibliotheque}

\end{document}